\documentclass[twocolumns]{aa}
\usepackage{graphicx}

\usepackage{epsfig}
\usepackage{textcomp}

\def\mum{\textmu m}

\begin{document}

\title{COSMIC EVOLUTION OF THE GALAXY MASS AND LUMINOSITY FUNCTIONS BY 
MORPHOLOGICAL TYPE FROM MULTI-WAVELENGTH DATA IN THE CDF-SOUTH} 

\author{A. Franceschini\inst{1}, 
  G. Rodighiero\inst{1}, P. Cassata\inst{1}, S. Berta\inst{1},
  M. Vaccari\inst{2},  M. Nonino\inst{3}, E. Vanzella\inst{3}, E. Hatziminaoglou\inst{4}, 
  J. Antichi\inst{1}, S. Cristiani\inst{3} }

   \offprints{A. Franceschini}

   \institute{
   Dipartimento di Astronomia, Universit\'a di Padova, Vicolo Osservatorio 5, 35122 Padova, Italy
         \email{franceschini@pd.astro.it}
         \and
   Astrophysics Group, Blackett Laboratory, Imperial College, Prince
   Consort Road, SW7 2AZ, London, UK
         \and
   INAF/Osservatorio di Trieste, Via Tiepolo 11, 34131 Trieste, Italy
         \and
  Institute de Astrofisica de Canarias, C/ Via Lactea s/n, E-38200 La Laguna, Spain}

   \date{Received date; accepted date}


 \abstract{
We constrain the evolution of the galaxy mass and luminosity functions from
the analysis of (public) multi-wavelength data in the Chandra Deep Field South
(CDFS) area, obtained from the GOODS and other projects,
and including very deep high-resolution 
imaging by HST/ACS. Our reference catalogue of faint high-redshift galaxies,
which we have thoroughly tested for completeness and reliability, comes from
a deep ($S_{3.6}\geq 1\ \mu$Jy) image by IRAC on the Spitzer Observatory.
These imaging data in the field are complemented with extensive optical
spectroscopy by the ESO VLT/FORS2 and VIMOS spectrographs, while
deep K-band VLT/ISAAC imaging is also used to derive further complementary 
statistical constraints and to assist the source identification 
and SED analysis.    We have selected a highly reliable
IRAC 3.6$\mu$m sub-sample of 1478 galaxies with $S_{3.6}\geq 10\ \mu$Jy, 
47\% of which have spectroscopic redshift, while for the remaining objects 
both COMBO-17 and $Hyperz$ are used to estimate the photometric redshift. 
This very extensive dataset is exploited to assess evolutionary effects in
the galaxy luminosity and stellar mass functions, while luminosity/density evolution 
is further constrained with the number counts and redshift distributions.
The deep ACS imaging allows us to differentiate these evolutionary paths by 
morphological type, which our simulations show to be reliable at least up 
to $z\sim 1.5$ for the two main early- (E/S0) and late-type (Sp/Irr) classes.

These data, as well as our direct estimate of the stellar
mass function above $M_\ast h^2=10^{10} M_\odot$ for the spheroidal subclass, 
consistently evidence a progressive dearth of such objects to occur
starting at $z\sim 0.7$, paralleled by an increase in luminosity.
A similar trend, with a more modest decrease of the mass function,
is also shared by spiral galaxies, while the irregulars/mergers
show an increased incidence at higher z.
Remarkably, this decrease of the comoving density with redshift of the
total population appears
to depend on galaxy mass, being stronger for moderate-mass, but almost
absent until $z=1.4$ for high-mass galaxies, thus confirming 
previous evidence for a "downsizing" effect in galaxy formation.
%
%
Our favoured interpretation of the evolutionary 
trends for the two galaxy categories is that of a progressive
morphological transformation (due to gas exhaustion and, likely, merging) 
from the star-forming to the passively evolving phase, starting
at $z\geq 2$ and keeping on down to $z\sim 0.7$. The rate of this process 
appears to depend on galaxy mass, being already largely settled 
by $z\sim 1$ for the most massive systems.


}
 \authorrunning{A. Franceschini, G. Rodighiero, P. Cassata et al.}
 \titlerunning{Evolution of the Galaxy Mass and Luminosity Functions}

\maketitle

\section{INTRODUCTION}
\label{intro}

Subject of active, as much as inconclusive, investigation during the
last 40 years or so, the cosmological origin of the Hubble galaxy 
morphological sequence can now be very effectively constrained
by combining the unique imaging capabilities of HST/ACS with 
the wide IR multi-wavelength coverage offered by the Spitzer Space Telescope
and the remarkable photon-collecting power and multiplexing 
of spectrographs on large ground-based telescopes (VLT, Keck).        
As much a complex process as it might 
have been -- involving both gravity and hydrodynamics
(see e.g. Baugh et al. 2005), and possibly other physical processes such
as black-hole formation and accretion, tidal interactions and merging, and 
feedback from stellar and nuclear activity (Springel et al. 2005) 
-- we have now a definite chance to observe it in operation.

At the current stage, however, the subject remains still rather controversial.
While slow infall of primordial gas may explain disk formation in a relatively simple 
way, (e.g. Mo, Mao \& White 1998), we still lack an adequate understanding 
of spheroid formation. On one side, the homogeneity of the early-type population
and tightness of the fundamental plane might suggest that these galaxies have 
formed from a single 
monolithic collapse, an early aggregation of lumps of gas turning into stars in 
the remote past ($z_{form} \ge$3) via a huge burst-like episode followed by quiescence
(Eggen et al. 1962; Larson et al. 1975; Chiosi and Carraro 2002).

This however is in apparent contradiction with recently favoured models of hierarchical 
galaxy formation which postulate that early-type galaxies are assembled at later
times by stochastic merging of lower-mass galaxies, either accompanied 
by strong star formation activity (e.g. White et al. 1978;
White et al. 1991; Somerville \& Primack 1999; Cole et al. 2000), or through
more "silent" dry-merging and dynamical aggregation (Bell et al. 2005a,
Tran et al. 2005).
In such case, ellipticals would be formed over timescales comparable to the 
Hubble time, with a major fraction of the mass assembly taking place around 
$z \sim  1$ (e.g. Somerville et al. 2001), and virtually all massive 
galaxies disappearing by $z \geq 1.5$. 
Benson et al. (2002) find popular hierarchical models to produce as many 
spheroids with highly inhomogeneous colors as observed locally, but to
underpredict the proportion of homogeneous, passive objects at redshifts $z \sim 1$. 
This suggests that while the star formation rate in spheroidals at low redshifts 
($z \le 1$) 
is predicted correctly, the formation rate at higher redshifts is underestimated. 
On the other hand, recent results from the K20 project (see Daddi et al. 2004a 
and references therein) claim that semianalytic models underestimate 
the number of massive galaxies at $z \sim 2$ 
by about a factor of 30 and suggest that the assembly of 
massive galaxies took place at substantially earlier epochs
than predicted by these models. 

Observational constraints on the star formation history have been inferred from
the broad-band colors, line strength indices and stellar chemical abundances. 
When referred to massive ellipticals, these observations often suggest that the bulk of 
stars might have been formed in a remote past. However, some secondary activity of 
star formation in the recent past is also evident: nearby ellipticals show a large 
variety of morphological and kinematical peculiarities (e.g. Longhetti
et al. 2000)
and a considerable spread of stellar ages, particularly for the field population 
(Thomas et al. 2005). Strong evolution in the population of early-type galaxies has 
been reported by Kauffmann, Charlot \& White (1996) and Kauffmann \& Charlot (1998),
which has been considered to support the hierarchical galaxy formation models.

Published results from high redshift galaxy surveys appear not unfrequently
in disagreement with each other, and conflicting conclusions are 
reported (see Faber et al. 2005 for a recent review about galaxy evolution 
at $z<1$). This is partly due to the small sampled areas and the 
corresponding substantial field-to-field variance. 
However, a more general problem stems from the apparent conflict between reports
of the detection of massive galaxies at very high redshifts (e.g. Cimatti et al.
2004; Glazebrook et al. 2004; Labb\'e et al. 2005; Daddi et al. 2005a,b) 
and indications for a fast decline in the in the comoving number density at $z>1$ 
(Franceschini et al. 1998; Fontana et al. 2004).

In summary, whereas early monolithic collapse and hierarchical models imply radically 
different histories for spheroids, neither the theoretical predictions nor the 
observational constraints for field galaxies have yet been sufficiently definitive for 
precise conclusions to be drawn in favour of one or the other scenario.

The most direct way of constraining the evolutionary history of galaxies and trying
to resolve the discrepancies would
be to derive the redshift-dependent luminosity and mass functions from
deep unbiassed surveys.  This has been pursued by a number of teams,
relying on either U-band optical selection as a probe of SFR density 
(Lilly et al. 1996; Madau et al. 1996) and UV selection by GALEX
(Schiminovich et al. 2005), or observations in the K-band
(Cowie et al. 1996; Cimatti et al. 2002; Dickinson et al. 2003; Franx et al. 2003; 
Drory et al. 2004; Fontana et al. 2004; Bundy, Ellis \& Conselice 2005).

We contribute to this effort by exploiting in this paper very deep public 
imaging by the IRAC photometric camera on the Spitzer Observatory to 
select a most unbiased sample of  high-z ($z\leq 2$) near-IR galaxies. 
We use for this the IRAC Channel-1 3.6$\mu$m data
over 160 arcmin$^2$ in the Chandra Deep Field South (CDFS) 
taken within the GOODS project (Dickinson et al. 2004).
The other IRAC imaging data in the field are either redundant (Channel-2 at 
4.5$\mu$m, too close to channel-1 and somewhat less sensitive), or
include non-stellar contributions by the galaxy ISM (the longer wavelength 
Channels-3 and 4) which would far complicate the interpretation.

Near-IR surveys are best suited for the study of faint high-redshift galaxy 
populations, for various reasons.  Compared to UV-optical selection, 
the observed fluxes are minimally affected by dust extinction. At the same 
time they are good indicators of the stellar mass content of galaxies 
(Dickinson et al. 2003; Berta et al. 2004), and  closer to provide a 
mass-selection tool.  
For typical spectra of evolved galaxies, the IRAC Channel-1 3.6$\mu$m also 
benefits by a K-correction particularly favourable for the detection of
high-redshift galaxies, particularly if we consider that the $H^-$ opacity
minimum (corresponding to a typical peak in galaxy's SEDs) 
at $\lambda\simeq 1.7\ \mu$m in stellar athmospheres (Simpson \& Eisenhardt 1999 ) 
falls within the waveband of Channel-1 for z=1 objects.

In addition to the Spitzer observations, the GOODS project 
has provided the community with an unprecedented amount of high quality
optical and near-IR 
data in CDFS, particularly the very deep 4-band ACS imaging, allowing the most
accurate morphological analysis currently possible over an appreciable area.
Building on our previous experience of faint galaxy imaging and statistical 
modelling (Rodighiero et al. 2001; Cassata et al. 2005), on our tools for 
spectro-photometric analysis (Poggianti, Bressan \& Franceschini 2001;
Berta et al. 2004), and expertises for reduction of deep IR imaging data 
from space (e.g. Rodighiero et al. 2004 for ISO data; Lonsdale et al. 2004, 
Hatziminaoglou et al. 2005, Rodighiero et al. 2005 for Spitzer data), 
we illustrate in this paper the power of combining 
such multi-wavelength information in the analysis of the evolutionary
mass and luminosity functions of faint galaxies.

Since the spectroscopic follow-up is currently only partial in the field, 
and to avoid as far as possible confusion problems in the IRAC data,
we limit our analysis to moderate depths. In spite of this,
the constraints on the history of massive galaxy evolution are already relevant.
Pushing the analysis to the IRAC sensitivity limits will allow to extend our 
conclusions further down in mass/luminosity and up in $z$.

 Though substantially larger than HDF's, our survey field is still of 
moderate size. Eventually, a complete understanding of the influence
of large-scale structure on the evolutionary history and to reduce the effects of
cosmic variance will need new-generation datasets on large areas 
(e.g. COSMOS, Scoville et al. 2004).    

The paper is structured as follows. Section \ref{data} describes the 
optical and IR (Spitzer) data, the near-IR data and ACS/HST imaging, the spectroscopic
data used in our analysis, and our criteria for catalogue combination and
merging. Section \ref{morph} details on our quantitative morphological
analysis of faint galaxies, and Section \ref{photoz} summarizes our effort
for the photometric redshift estimate. 
 Our statistical analyses are reported 
in Section \ref{stati}, while Section \ref{mf} is dedicated to derive 
the mass function in stars of our galaxies. 
A comparison of observed and model number counts is discussed
in Section \ref{model}. Sections \ref{discussion} and \ref{conclusion}
summarize our results and conclusions.  

We adopt in the following a standard set of values for the cosmological
parameters $\Omega_M$=0.3 and $\Omega_{\Lambda}=0.7$, while for ease of
comparison with other published results we express the dependence on the
Hubble constant in terms of the parameter $h\equiv H_0/100\ Km/s/Mpc$
and provide the relevant scaling factors.

\section{Observations and Data Analysis}\label{data}

\subsection{Deep IR Imaging with Spitzer }
 
The GOODS southern field is located in the Chandra Deep Field South 
(CDFS) area, that is centered at RA(J2000)= 03:32:30.37 and DEC(J2000)=
-27:48:16.8. 
   The IRAC Spitzer observations in the field include deep
imaging in four near-IR bands (3.6, 4.5, 5.8 and 8.0 $\mu$m). 
The exposure time per channel per sky pointing was 23 hours as a minimum.
Due to the fact that observations were done in two epochs, it is approximately 
two times this in a central deeper strip where the two overlap.
We exploit in this paper a galaxy catalogue that we have derived from
the 3.6 $\mu$m IRAC data.        

We started the data reduction using products generated
by the Spitzer Science Center (SSC) Basic Calibrated Data (BCD) pipeline. 
We used all BCDs available in the archive at the end of November 2004.
The archived data were processed with pipeline version S10.5.0 provided by
the Spitzer Science Center.
We applied an additive correction factor to each BCD frame in order to remove 
the median background. 
We have then processed and mosaiced all corrected BCDs  
within the Mopex 
package{\footnote{
 Mopex performs the processing and mosaicing of both
IRAC and MIPS Spitzer imaging data. Details can be found in
http://ssc.spitzer.caltech.edu/postbcd}}, 
using a standard procedure that accounts for cosmic radiation hits, 
outliers, flat-field and distortions of the detector.
The pixel size in the final map is 0.6 arcsec/pixel.
The GOODS IRAC mosaic with a significant sky coverage 
(sky pixel repetition factor$>20$) covers an area of approximately $12\times 18$
square arcminutes on the sky.

The IRAC source extraction has been performed with SExtractor (Bertin \& Arnouts, 
1996).
Assuming that essentially all the sample sources are seen as point-like by the
IRAC $\sim 3\ arcsec$ imager, we computed the fluxes within a 
5.9 arcsec diameter aperture and applied a correction factor derived from 
the stars in the IRAC images to assess the total fluxes.
In the case of extended sources, we used Kron like magnitudes 
(AUTO$\_$MAG output parameter in SExtractor).

\subsection{Near-IR Ground-based Imaging }
\label{Kimag}

As part of GOODS, near-infrared imaging observations of the CDFS have being carried out
in J, H, Ks bands, using the ISAAC instrument mounted on the ESO VLT
telescope. We made use of the publicly available J and K band imaging (version 1.0, 
released{\footnote {http://www.eso.org/science/goods/releases/20040430/}} by 
the ESO/GOODS team in April 2004).
This data release includes 21 fully reduced VLT/ISAAC fields in J and
Ks bands, covering 130 arcmin$^2$ of the GOODS/CDFS region. It also
includes mosaics of the co-adjoined tiles as single FITS files in J
and Ks bands, as well as the corresponding weight-maps.

To provide a homogeneous photometric calibration across the entire field,
the GOODS team at ESO have rescaled all images to the same zero point
(26.0 in the AB system). The final mosaics have a pixel scale of 0.15".
We have run SExtractor on the J and K mosaics to 
obtain total magnitudes (BEST$\_$MAG output parameter in SExtractor) for all the 
objects in the field.

\subsection{The Near-IR Source Selection Functions }
\label{compl}

The completeness of our 3.6 $\mu$m catalogue was assessed through 
numerical simulations. A number of sources spanning the 0.3-30 $\mu$Jy 
flux range was added to the image at random positions using a synthetic 
TinyTim PSF (Krist, 2002). A conservative overall figure of 1600 sources (corresponding 
to 200 beams/source) was chosen as the additional source density in
order to avoid confusion effects.
  
\begin{figure}
\centerline{
\psfig{file=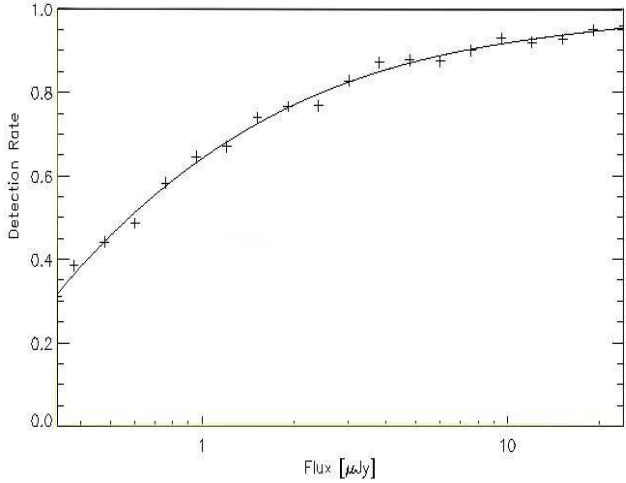,width=9cm,height=8cm}
}
\caption{Selection function of the 3.6 $\mu$m band Spitzer sample in the CDFS,
as from our Monte Carlo simulations. See text for details.}
\label{dr36}
\end{figure}

Source extraction was carried out as done on the original image, and 
inputs and outputs were cross-correlated using a 1.5 pixel search radius 
(corresponding to 0.9 arcsec). The results are shown in Figure
\ref{dr36}, where a fit to the selection function curve is also shown.
Our sample turns out to be $\sim60$\% complete above 1 $\mu$Jy, $\sim90$\% 
complete above 5 $\mu$Jy, and more than $\sim95$\% above 10 $\mu$Jy.
Further tests of the validity of our IRAC source selection will be
discussed in Sect. \ref{counts_obs} based on a comparison of our and 
independent results on the galaxy number counts.

The IRAC 3.6$\mu$m sample at 1 (10) $\mu$Jy includes 5622 (1646) 
sources, respectively.
 
We have also performed extensive Monte-Carlo simulations, including
the insertion of several IRAC/GOODS-like toy galaxies in the real image, to determine 
the limit of completeness in the K band. We applied the same extraction criteria 
used on the ISAAC images and we derived the detection rate as a function
of the simulated magnitude. We estimated the K-band sample to be more
than 90\% complete at $K<21$ (Vega reference system).

\subsection{ACS/HST Optical Imaging}

The core of the GOODS project was the acquisition and reduction of 
high-resolution HST/ACS imaging data obtained as an HST Treasury Program
(Giavalisco et al. 2004).
The GOODS ACS/HST Treasury Program has surveyed two separate fields 
(the CDFS and the Hubble Deep Field North) with four broad-band filters: 
F435W (B); F606W(V); F775W(i); and F850LP(z). Observations in the V, i 
and z filters have been split into five epochs, separated by about 45 degrees,
in order to detect transient objects. Observations in the B band are
taken during epoch 1 for both fields. 
Images taken at consecutive epochs have position angles increasing by
45$^\circ$. The total exposure times are 2.5, 2.5, 5 orbits in the V, i and z bands,
respectively. The exposure time in the B band is three orbits. In August 2003 the
GOODS team released version 1.0 of the reduced, stacked and mosaiced images for
all the data acquired over the five epochs of observation. To improve the
point spread function (PSF) sampling, the original images, which had a scale of
0.05 arcsec pixel$^{-1}$, have been drizzled on to images with a scale
of 0.03 arcsec pixel$^{-1}$.
We used the version 1.0 of the image catalogue. 

The data set is complemented with the ACS/HST catalogs released by the
HST/GOODS team in October 2004, containing the photometry in B, V, $i$
and $z$ bands. The source extraction and the photometric measurements 
have been performed by the GOODS team running a modified version of SExtractor
on the version 1.0 images. Particular attention has been payed to the photometry 
of faint sources, thanks to the careful determination of the local sky background.
As for the case of the K-band imaging, we have considered the total magnitudes
(as from the BEST$\_$MAG output parameter in SExtractor).
Moreover, for the purposes of a detailed morphological analysis, 
we used the high resolution deep imaging
carried out by ACS/HST in the z-band (version 1.0, Giavalisco et al. 2004).

\subsection{Optical Spectroscopy }
\label{spec}

In the last few years ESO has performed various systematic spectroscopic
observational programs in the CDFS area by using all available 
multi-object spectrographs (i.e. FORS1, FORS2, VIMOS; see Cimatti et al. 
2002b; Le Fevre et al. 2004a,b; Szokoly et al. 2004; Vanzella et al. 2005).
The contributions of the major projects are hereby summarized.

The VIMOS VLT Deep Survey (VVDS, Le Fevre et al. 2004a) observed a
large sample  of galaxies in the CDFS area. The redshift data have 
been released to the community{\footnote {http://cencosw.oamp.fr}}: a total
of 1599 objects with  $I_{AB} \leq 24$ have measured redshifts. The
completeness in redshift measurement for the targeted objects is
high, better than 84\%, and 784 of these fall within the
GOODS-South ACS field. 

In the framework of the GOODS project, a large sample of galaxies in 
the CDFS has been spectroscopically
targeted (Vanzella et al. 2005). A total of 303 objects with $z_{850}
\la 25.5$ have been observed with the FORS2 spectrograph, providing
234 redshift determinations. 
The reduced spectra and the derived redshifts are released to the
community{\footnote {http://www.eso.org/science/goods/}}. 

Further GOODS spectroscopic observations have been carried out with 
FORS2 at VLT (Szeifert et al., 1998) using the 300I grism, without
blocking filter (ESO programs 170.A-0788 and 074.A-0709). 
This configuration gives a low resolution ($R\sim 850$) in the
wavelength range $\sim 5600-10000\AA$. Data reduction has been
detailed in Vanzella et al. (2005).

Another fraction of spectroscopic redshifts has been obtained as part of the K20 
survey (Cimatti et al. 2002b), which have also been recently made public{\footnote 
{http://www.arcetri.astro.it/$\sim$k20/spe\_release\_dec04/index.html}}.
The K20 sample includes 546 objects to $K_{\rm s} \leq 20$ (Vega
system)  over two independent fields (52 $arcmin^2$ in total, 32
$arcmin^2$ in the CDFS). The spectroscopic redshift
completeness is  94\% and 87\% for $K_{\rm s} \leq 19$ and $K_{\rm s}
\leq 20$ respectively.

All the spectroscopic redshifts available in the version 1.0
of the GOODS/CDFS ACS catalogue have been compiled in a master list{\footnote
{http://www.eso.org/science/goods/spectroscopy/CDFS\_Mastercat}}.

%
%

\subsection{Infrared to Optical Source Associations}
\label{assoc}

The excellent quality of HST imaging allowed us to obtain reliable associations
for most of the Spitzer IR sources with the corresponding optical counterparts.
On the other hand, the much worse spatial resolution of IRAC compared to ACS
implies some significant complication in source association, which is
discussed below.

It should be noticed that the deep $K$-band ISAAC images might in principle
provide decisive support for the identification. However, only a fraction 
($\sim 60\%$, see Sect. \ref{multicat})
of the IRAC Spitzer 3.6 $\mu$m sources turned out to have counterparts in the K-band
(partly because the ISAAC imaging covered only $\sim$80\% of the Spitzer/ACS 
common area). For this reason, our identification of the Spitzer sources
followed a two-step procedure.

We have first looked for optical identifications by matching with the ACS 
$z$-band catalogue. This provided us with reliable associations for a large
majority of the 3.6 $\mu$m objects. 
We have then proceeded to match all Spitzer sources with sources extracted
from the ISAAC $K$-band images. This second step was needed not only to
obtain near-IR data essential for the SED fitting, photometric redshifts, etc.,
but also to disentangle the correct identification for 
dubious matches, multiple associations, or confused sources emerging from
step one.

\subsubsection{Associations with the ACS $z$-band Data}
\label{assocz}

The validity of the association of the 3.6 $\mu$m with $z$-band ACS 
sources was verified here by using the likelihood ratio
technique introduced by Sutherland \& Saunders (1992).
We adopt here the formalism described by Ciliegi et al. (2003).

The likelihood ratio (LR) is defined as the ratio between the
probability that a given source at the observed position and with the measured
magnitude is the true optical
counterpart, and the probability that the same source is a chance unrelated object:
\begin{equation}
LR = q(m)f(r)/n(m)
\label{e}
\end{equation}
where $n(m)$ is the surface density of objects with magnitude $m$;
$q(m)$ is the expected distribution as a function of magnitude of the
optical counterpart; $f(r)$ is the probability distribution function
of the positional errors.

In the presence of more than one counterpart in the errorbox, the
reliability $Rel_j$ for object $j$ being the correct identification is:
\begin{equation}
\label{rel}
Rel_j = \frac{(LR)_j}{ \sum_i(LR)_i + (1 - Q)}
\end{equation}
where the sum is over the set of all candidates for this
particular source and $Q$ is the probability that the optical
counterpart of the source is brighter than the magnitude
limit of the optical catalogue ($Q = \int^{m_{lim}} q(m) dm$).

For each IRAC source we adopted a mean positional error of 1 arcsec, and we
assumed a value of 0.1 arcsec as the
optical positional uncertainty.
We choose a search radius of 1.5 arcsec from the position of the centroid
of the infrared error box to look for the possible optical counterparts.

\begin{figure}
\centerline{
\psfig{file=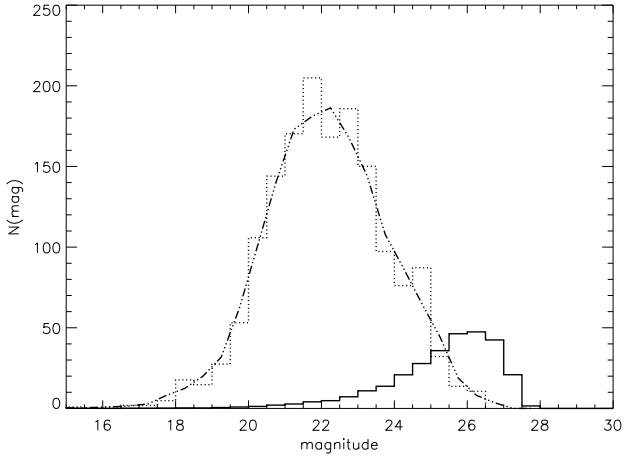,width=6cm,angle=90}
}
\caption{  AB magnitude distributions of background objects (solid line)
  and 'real' detections (real(m), dotted line) estimated from the
  optical objects detected in the $z$ band within a radius of 1.5
  arcsec around each infrared sources. The smooth curve fitted to the
  real(m) distribution (dot dot dot dashed line) has been used as input in the
likelihood calculation.}
\label{like}
\end{figure}

Figure \ref{like} shows the resulting magnitude distribution of the correct 
identifications (dotted histogram) together with the expected 
distribution of background objects unrelated to the infrared sources
(solid histogram).
The smooth curve fitted to the former (dot dot dot dashed line)
has been used as input in the likelihood calculation.  
Since the fraction of infrared sources with more than one possible optical
counterpart is $\sim$15\% (see below), this corresponds to an expected fraction of
correct identifications above the magnitude limit of the optical
catalogue of the order of $\sim$85\%. On this basis we adopted for the $Q$ 
parameter the value $Q = 0.85$
(to check how this assumption could affect our results, we
repeated the analysis using different $Q$ values 
in the range 0.5-1.0: no substantial difference in the final number
of identifications and in the associated reliability has been found).

Once $q(m)$, $f(r)$ and $n(m)$ have been determined, we computed the
LR value for all the optical sources within a distance of 1.5 arcsec
from the infrared positions.
The best threshold value for LR ($L_{\rm th}$) is then to be defined,
in order to discriminate between spurious and real identifications.
As discussed by Ciliegi et al. (2003),  $L_{\rm th}$
has to be small enough to avoid missing many real
identifications, and be large enough to keep the number of spurious
identifications as low as possible.

For the LR threshold we adopted the value $L_{\rm th}~=0.15$. With this 
and from eq. \ref{rel} with $Q=0.85$,
all the optical counterparts of infrared sources with only one
identification (the majority in our sample) and  $LR>LR_{\rm th}$ have
a reliability greater than 0.5.

The LR analysis has been performed for the whole IRAC sample to 
$S_{3.6}=1\ \mu$Jy.

In order to minimize the problem of misidentification,
we have checked by visual inspection all the optical associations of each
IR source with $S_{3.6\mu m}>10 \mu$Jy, our highly reliable and
complete sub-sample.
For stars and isolated objects, the cross-correlation is unambiguous:
the shape and the peak of the infrared contours overlayed on the $z$ image
have confirmed the associations. However, for several IRAC sources, 
particularly for the extended ones, the relationship between 
infrared and optical emissions may be much more uncertain: few optical
sources can lie inside the IRAC errorbox and contribute to the infrared 
emission. Such cases are dealt with in the following subsections.

\subsubsection{Multiple Associations and Confused Sources}

\begin{figure*}
\centerline{
\psfig{file=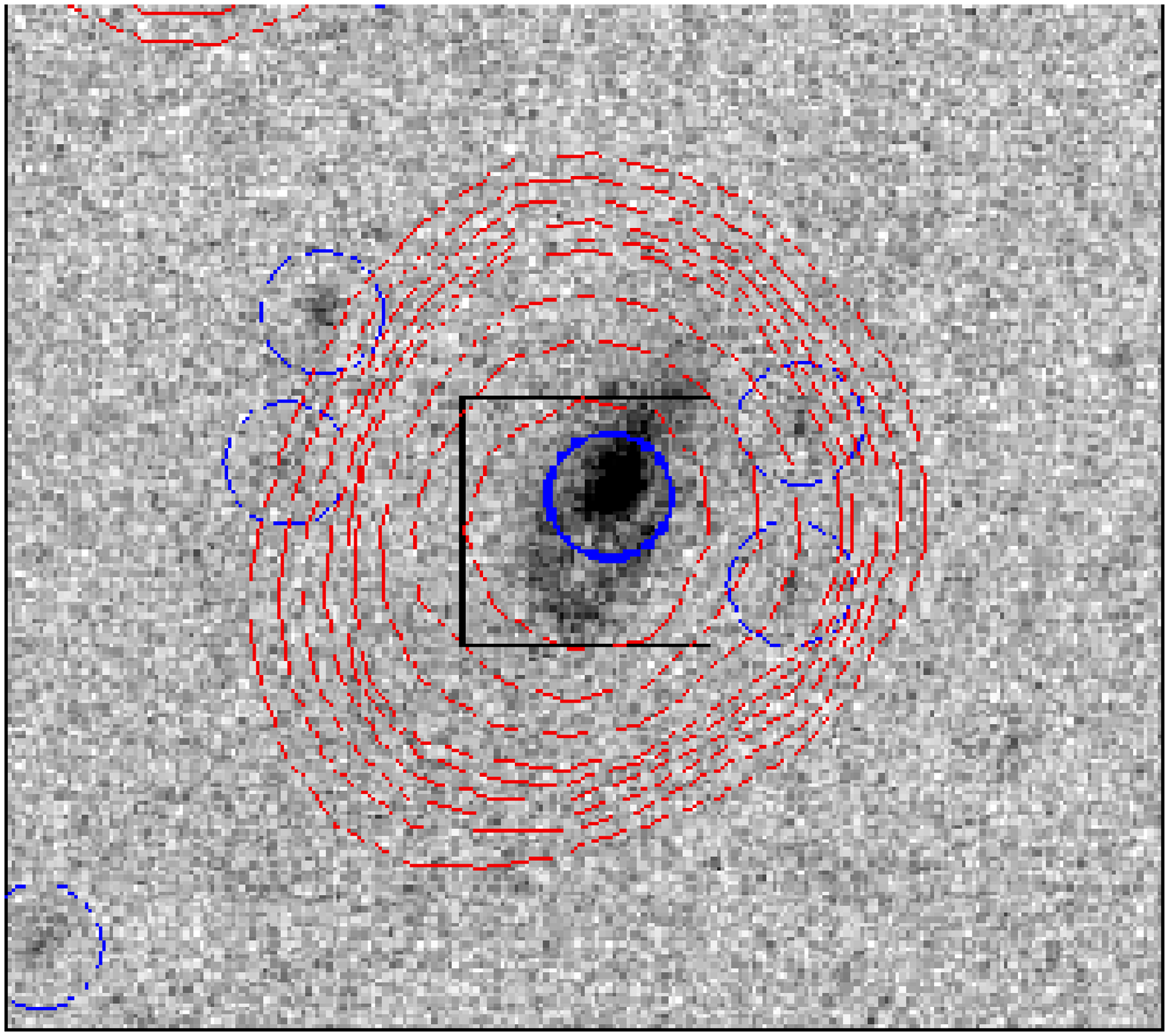,width=5.7cm}
\psfig{file=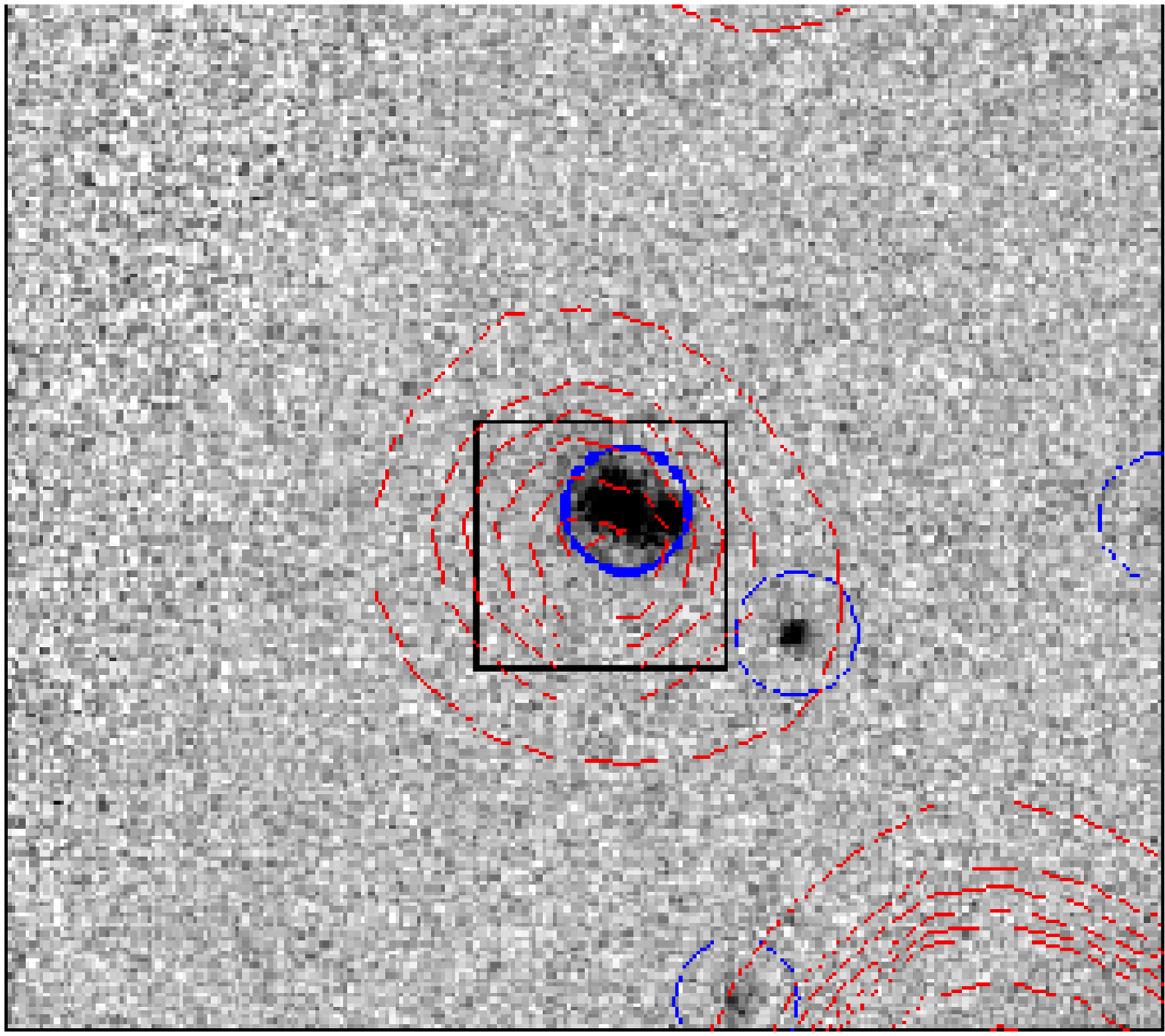,width=5.7cm}}
\caption{Examples of IRAC sources with multiple optical counterparts.
For the two sources the 3.6 $\mu$m countours (red lines) are overlaid on 
a zoom of the optical $z$-band ACS image (9''$\times$8'' size). 
The black squares mark the positions of the
original IRAC centroids, while the blue circles denote the optical
counterparts. The thicker blue circles indicate our final
associations, choosen with the automatic procedure discussed in
Section \ref{assoc}.}
\label{cooo1}
\end{figure*}

We found that $\sim$10\% ($\sim$15\%) of the $S_{3.6\mu m}>10\ \mu$Jy sources
have more than one optical counterpart within 1 (1.5) arcsec from the IR 
centroid position.
Examples are shown in Figure \ref{cooo1}, where we report zooms on the 
optical $z$-band image with the 3.6 $\mu$m contours overlaid (red lines)
for a couple of sources. The black squares mark the positions of the
original IRAC catalogue, while the blue circles are the optical
counterparts. The thicker blue circles indicate our final
associations, choosen by the automatic procedure discussed above.

In many such cases, the LR analysis was able to disentangle the correct
identification:  in the lack of independent information (Sect. \ref{assocK})
we assumed that the object with the highest likelihood ratio value was the 
correct counterpart of the IRAC source.

\begin{figure*}
\centerline{
\psfig{file=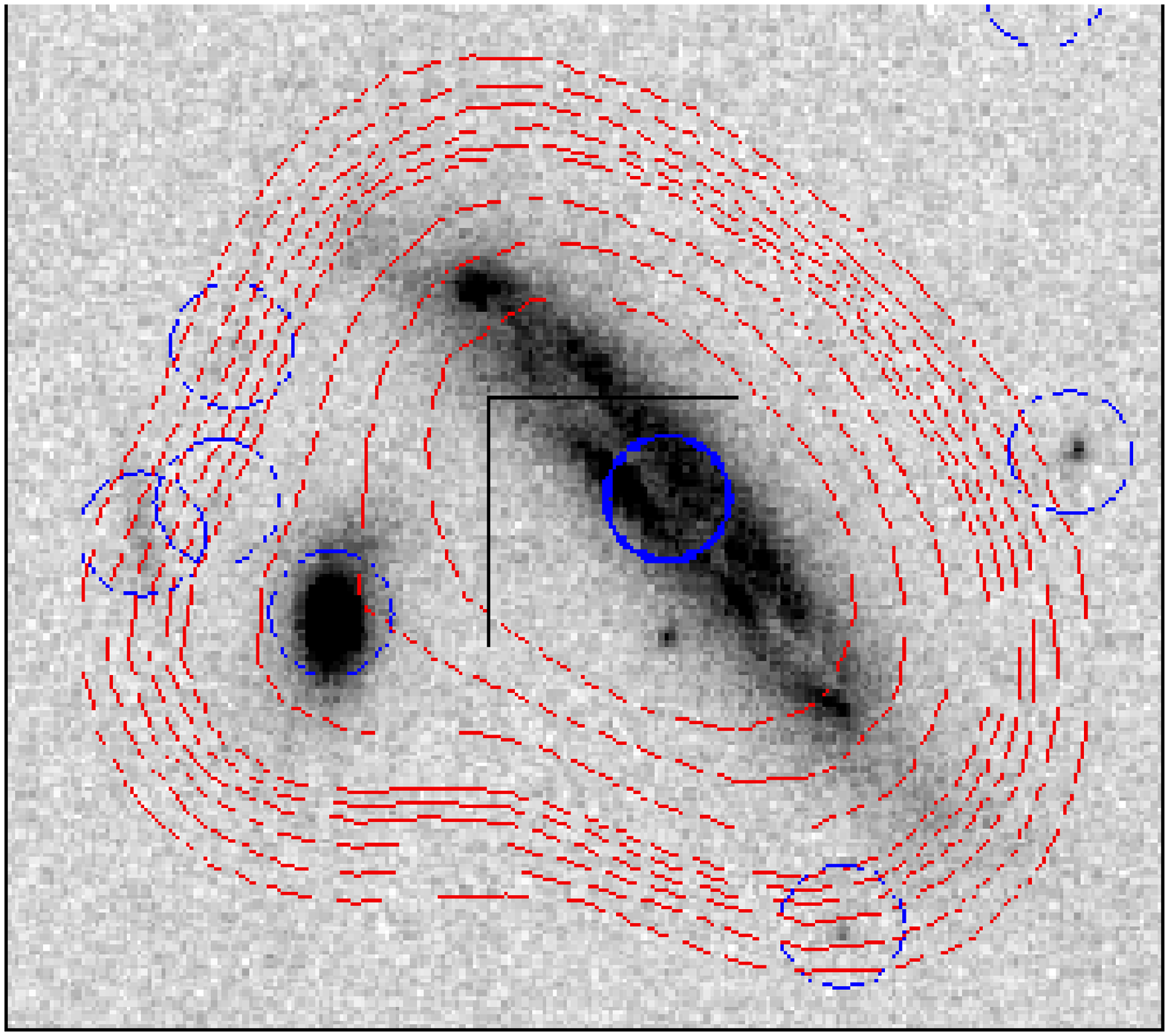,width=5.7cm}
\psfig{file=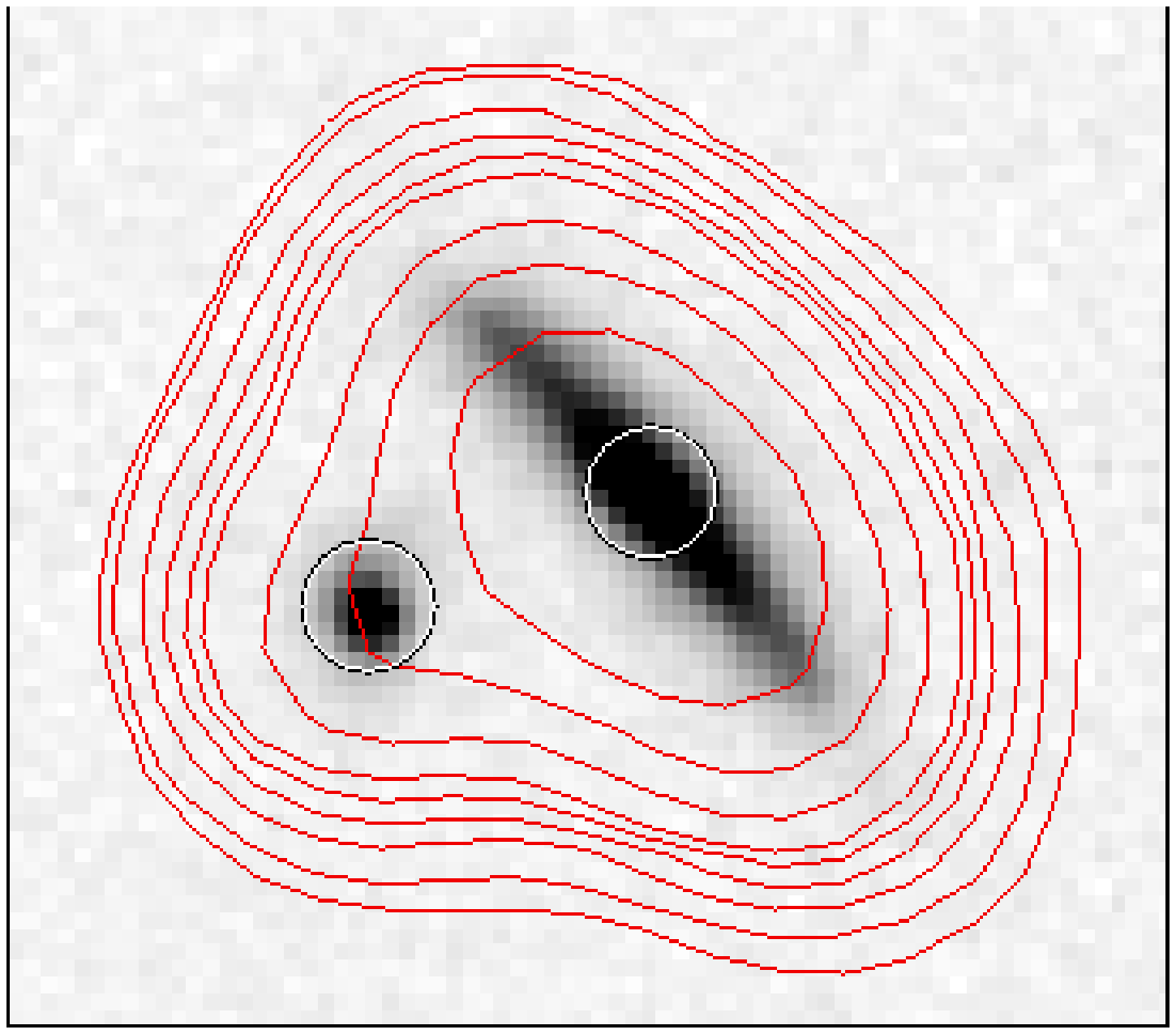,width=5.7cm} 
}
\caption{Example of a blended IRAC source. The meaning of the line types 
and the image size are the same as in Fig. \ref{cooo1}.  
The left panel reports an overlay of the IRAC 3.6$\mu$m contours
over the $z$-band ACS image. The two brighter optical sources, 
contributing to the IRAC flux, are at the same redshift ($z\sim 0.37$).
The right hand panel overlays the IRAC contours to the
ISAAC $K$-band image. The two components take 83\% and 17\%
of the $K$ flux, corresponding to $K_{AB}=18.72$ and 20.50.
This is adopted here as the flux ratio of the two sources 
assumed to produce the 3.6$\mu$m blend.}
\label{cooo2}
\end{figure*}

A fraction of the IR sources with multiple optical counterparts display clear
signs of confusion, due to the largely different spatial resolution of the Spitzer
and HST images. In such cases an extended 3.6 $\mu$m emission appears to emerge
from more than one optical source (generally by no more than two),
and there is no obvious indication of the level of contribution by the various
optical components to the IR flux. An example is given in Figure \ref{cooo2}:
in this case the two brighter optical sources are at the same redshift
($z\sim 0.37$).

The fraction of confused sources in our matched IRAC/ACS sample above
$S_{3.6\mu m}=10\ \mu$Jy is of the order of $\sim$5\%.

Concerning the part of the sample with fluxes between 1 and 10 $\mu$Jy,
the fractions of sources with multiple optical associations (hence with
potential confusion problems) is 16\% for an association radius of
1 arcsec and 28\% for 1.5 arcsec radius.  For these we used the LR 
value to identify the correct optical counterpart.

\subsubsection{Matching with the ISAAC $K$-band Imaging}
\label{assocK}

To reduce the level of contamination and confusion in our samples
we have made full use of the deep ISAAC $K$-band imaging, which is
close enough to the 3.6 $\mu$m selection wavelength and at the
same time provides sub-arcsec resolution.

We have cross-correlated the $z$-band ACS catalogue with the ISAAC
image in the 130 sq. arcmin common area, by applying the LR
analysis presented in Sect. \ref{assocz}. This comparison allowed us to 
resolve the majority of the confusion problems in the complete 10 $\mu$Jy
sample. Our procedure was to split the 3.6 $\mu$m confused source
into as many components as revealed by the ISAAC image (typically
two), and to assign to each one a flux according to the observed 
ratios of the fluxes in the $K$-band. 
An example of the application of this procedure is reported in Fig. \ref{cooo2}.

In such a way we have resolved
60\% of the confused cases, and brought the fraction of confused
sources to only 2\% of the complete sample.

Due to the small number and negligible impact on our results of the 
latter, we have not attempted to apply complex deblending procedures 
to recover the 3.6 $\mu$m flux coming from each single components.
Instead, when building the observed SEDs for these 30 objects, we have
summed the optical fluxes of the components of the IRAC blend, 
at least where there was evidence for interaction from morphology
(see the example in Figure \ref{cooo2}).

In conclusion, our choice of a moderate flux limit 
($S_{3.6\mu m}=10\ \mu$Jy) for our reference complete sample
minimizes the photometric complications due to source confusion. 
For the fainter sample, used in the following only for extracting
morphology-selected number counts, the application of the procedure 
of using the ISAAC image did not bring a significant improvement 
in the statistics of multiple or confused sources.

\subsubsection{The Band-merged IRAC/GOODS Catalogue }
\label{multicat}

Once each 3.6 $\mu$m source has been assigned to its $z$-band -- and
whenever possible $K$-band -- counterparts, we proceeded by builiding 
a multi-wavelength catalogue, in order to include all the available
photometric bands. The cross-correlation with the ISAAC (J and K) and the
other HST bands was done using a search radius of 1.5 arcsec around
the $z$-band positions.
The area covered by HST observations, approximately $\sim 160$ square arcminutes,
is smaller than that originally covered by Spitzer, 81\% of which (130 sq. arcmin)
is also surveyed in $K$ by ISAAC.

Our final sample includes 1646 IRAC sources with $S_{3.6\mu m}>10\ \mu$Jy
(1478 turn out to be galaxies and 168 stars, see below)
and a total of 5622 sources with $S_{3.6\mu m}>1\ \mu$Jy inside the Spitzer/ACS 
common area of 160 sq. arcmin, 5302 of which are galaxies and 320 stars.
Sixty percent of the total sample sources have a $K$-band counterpart. 
This fraction reaches 75\% when considering the high-reliability limit 
of $S_{3.6\mu m}>10\mu$Jy.

\section{MORPHOLOGICAL CLASSIFICATION }
\label{morph}

We have performed a quantitative morphological classification using the $z$-band ACS 
images for all sample galaxies. For this we measured for each galaxy the set of 
parameters Concentration ($C$), Asymmetry ($A$) and clumpinesS ($S$) 
of the galaxy light distribution (CAS parameters).

The concentration roughly correlates with the bulge
to disk ratio, enabling distinction between bulge- and disk-
dominated galaxies (Abraham et al. 1996). The asymmetry differentiates
normal isolated galaxies from irregular/merging systems
(Conselice et al. 2000). The clumpiness (a measure of the uniformity
of the galaxy surface brigthness distribution) is expected to correlate 
with the ongoing star formation rate (Conselice 2003a).

\begin{figure}
\psfig{file=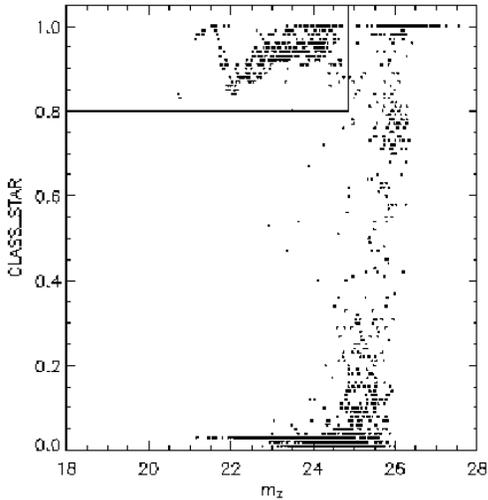,width=8.5cm} \hfil
\caption{Plot of the stellarity index ($CLASS\_STAR$) from the SeXtractor analysis
versus the $z_{AB}$ magnitude. Stars are selected in the rectangular box of high
stellarity and bright magnitude.  }
\label{star}
\end{figure}

When attempting a morphological classification of a large sample of
galaxies at widely different redshifts, it is crucial to take into
account the effects of the so-called morphological K-correction.
This considers that galaxies observed at bluer wavelengths 
tend to display a later morphological type.
For this reason we have used in this work the longest wavelength $z$-band 
(the F850LP ACS filter) to study the morphology of the entire sample: 
this band corresponds to about rest-frame I, V, B and U 
for galaxies at z=0.2, 0.5, 1 and 1.5 respectively.

However, as already discussed by various authors (Windhorst et al. 2002;
Papovich et al. 2003), this effect is
relevant mainly when comparing U rest-frame with visual (B or V rest-frame) 
morphologies. This means in our case that only some of the
galaxies at a redshift larger than $z\sim1.5$ may be significantly 
influenced by this effect.
Cassata et al. (2005, henceforth CA05) discuss in detail and try to quantify 
it by comparing the visual morphology of a sample of $z\sim1$ galaxies in 
their U and B rest-frame. They find that the morphology is preserved for all 
early-type galaxies, while some B rest-frame disks appear as irregular
in the U rest-frame (but this is not important in our approach, 
that segregates only early- from late-type galaxies). 
Given that the technique used for this work is calibrated on
the results by Cassata et al. (2005), the same conclusions of that 
work can be applied here.
Therefore, we do not expect our analysis to be affected by strong
morphological K-correction effects, especially considering that our main
points about the evolutionary mass and luminosity functions in the following 
Sects. concern the $z<1.5$ universe, and since we confine our analysis
to the basic early- to late-type structural differentiation.

Morphological classification based on the CAS parameter set is 
particularly effective in disentangling spheroid-dominated from
spiral/irregular galaxies.
Early type galaxies usually have a low value of asymmetry, an high
concentration and a small clumpiness. Cassata et al. (2005) find that early type
galaxies in the K20 catalogue (a subset of the current GOODS sample, 
Cimatti et al. 2002) 
occupy a rather precise domain of the CAS parameter space, as
measured on the ACS/GOODS images:
\begin{equation}
A < 0.2 \, \ ; \, C > 2.7 \; \ \, 0.0 < S < 0.3 \, ;  3A - S < 0.3 .
\label{eqCAS}
\end{equation}
Galaxies having CAS parameters lying in the region above are classified
as early-type, while the remaining are classified as late-type.
The stars are very effectively isolated by combining the $z$-band flux with
the stellarity index calculated by SExtractor ($CLASS\_STAR$) on the
$z$-band images.  Figure \ref{star} plots the stellarity index 
($CLASS\_STAR$) versus the $z_{AB}$ magnitude. Stars are easily identified 
in the rectangular box of high stellarity and bright magnitude.

We have first applied the above technique to the 1646 objects of our
3.6 $\mu$m catalog above 10 $\mu$Jy, and identified in such way 168 stars, 472
(32\%) early-type and 1006 (68\%) late-type galaxies in the complete sample.
This compares with a fraction of 26\% early-types (74\% spirals and irregulars) 
estimated by Bundy et al. (2005) among galaxies brighter than $z_{AB}=22.5$.

The morphology of a subsample of 155 galaxies (about 10\% of the objects 
in our 10 $\mu Jy$ catalog) has also been visually inspected, in order to 
check the reliability of our automatic technique.
We found that the visual and the CAS mophologies agree with each other
for 90\% of the 55 galaxies automatically classified as early-type and
for 97\% of those automatically classified as late-types. The
automatic classification procedure then provides a valuable and robust 
tool for disentangling early from late morphological type galaxies.

We have then addressed the more challenging problem of estimating structural 
properties for sources fainter than 10 $\mu$Jy and down to $S_{3.6}=1\ \mu$Jy.
To this end, we have applied our method to all 3976 sources detected within
this flux interval, of which 152 turned out to be stars, 548
early-type and 3276 late-type galaxies. 
Then the fraction of spheroidal galaxies to total drops from more than 30\%
in the bright sample to 14\% fainter than $S_{3.6}=10\ \mu$Jy.

It should be noted that, among all objects in this flux interval, $\sim$990
have z-band ACS magnitude fainter than 25 (927 of which classified by CAS
as late-type, 61 as early-type and 2 stars). Our simulation experiments 
(see also Cassata et al. 2005)  indicate that, beyond $z_{AB}\simeq 25$ mag,
the automatic classification is less reliable.

\begin{figure*}
\psfig{file=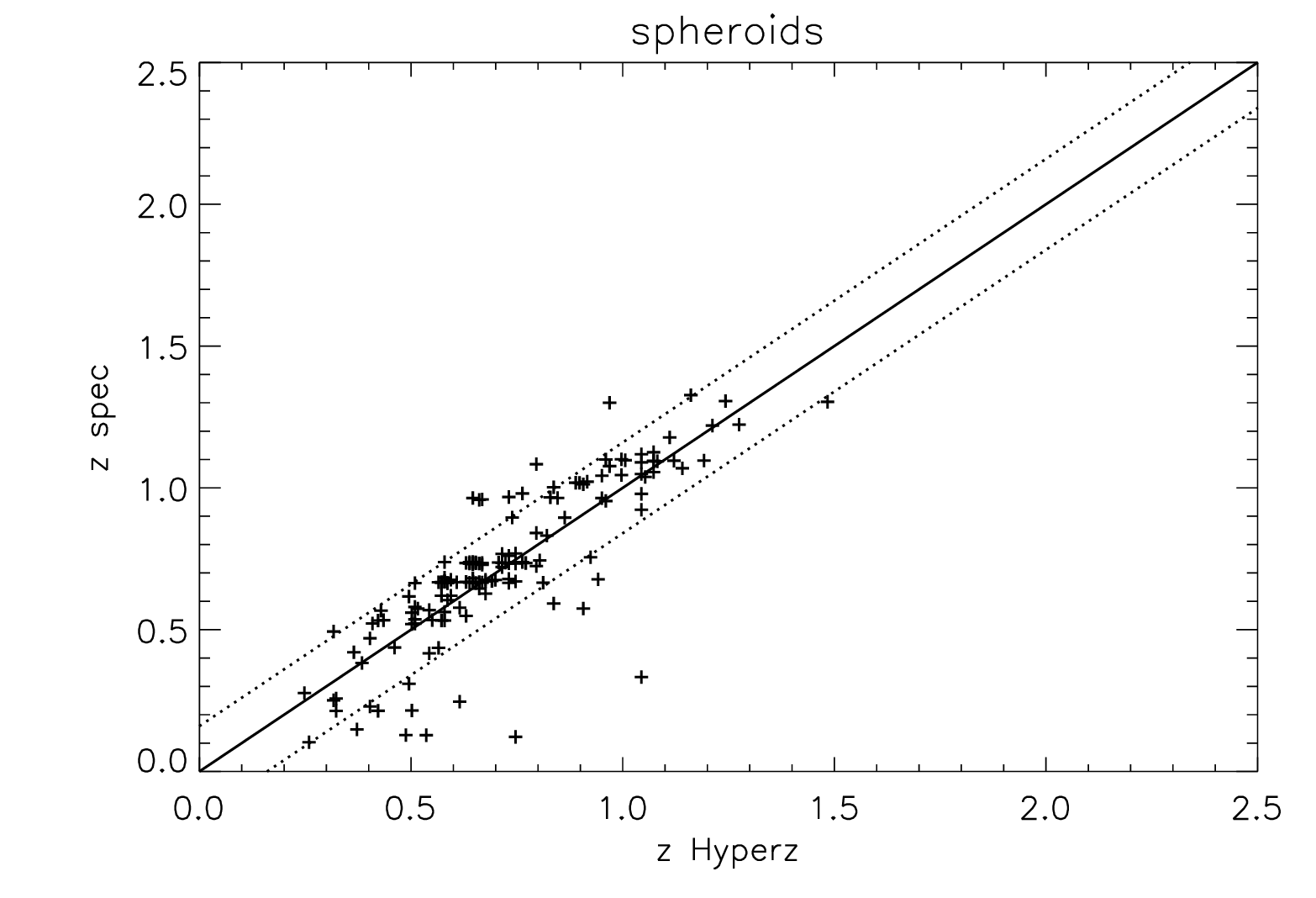,width=8cm} \hfil
\psfig{file=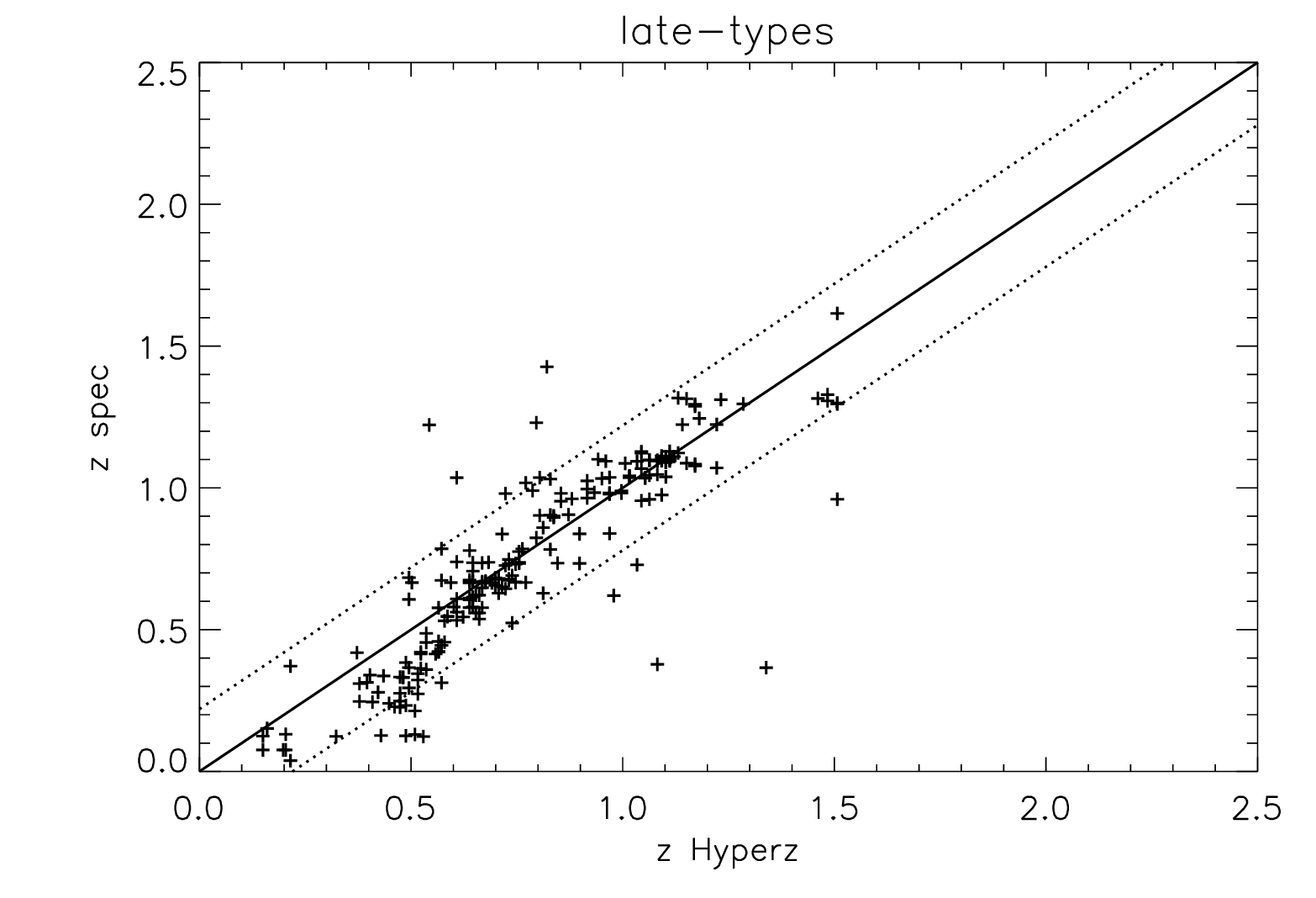,width=8cm}
\caption{Comparison of photometric versus spectroscopic redshifts for
  early-type (left panel) and late-type galaxies (right panel). We
  report only the sources with $S_{3.6}>$10 $\mu$Jy and for which $Hyperz$ found an highly reliable
  fit (corresponding $\chi^2$ probability above 99\%). The dotted lines delimit the 2$\sigma$ 
  uncertainty ranges.}
\label{zphot}
\end{figure*}

For this reason we have performed a visual check of a large subset
of the IRAC/GOODS galaxy sample, with particular attention to the 
990 optically fainter, in random magnitude order, to determine how many would be 
consistent with an early-type and how many with a late-type morphology.
As for the 548 CAS-classified ellipticals, visual inspection
shows that approximately 30\% may be consistent with 
having misclassified late-type morphologies. 
On the other hand, a CAS-classified late-type galaxy has a much lower
chance (of the order of 4\% at the 1 $\mu$Jy limit) to be a spheroidal system.

In conclusion, for spheroidal galaxies we evaluate that the overall uncertainty 
due to morphological misclassification of the faint optical counterparts is
such that their number ranges from 380 to 690 objects (i.e. from 10\% to 18\% of 
the total galaxy population) at the 1 $\mu$Jy sensitivity limit of our survey.
The corresponding relative uncertainty in the late-type morphological subset
is lower by a factor $\sim 5$, or of the order of 20\% at most.
This provides us with the upper and lower boundaries in the 
number counts differentiated by morphological class
(see Sect. \ref{counts_obs}).

\section{PHOTOMETRIC REDSHIFTS }
\label{photoz}

The spectroscopic data available in the GOODS/IRAC/GOODS area (see Section
\ref{spec}) have been cross-correlated with our $S(3.6_{\mu m})>10\mu$Jy 
complete catalogue, including 1478 extragalactic sources at this
flux limit. Of these, 695 (or 47\%) turned out to have a spectroscopically
confirmed counterpart.  

For an important fraction of the remaining galaxies, we have used public
photometric redshifts from COMBO-17 (Wolf et al. 2004), a multi-band
photometric survey entirely covering our field.
Wolf et al. have shown that the COMBO-17 photometric
redshifts are highly reliable and accurate for galaxies
with $R<24$ and $z< 1$, for which the typical 1-$\sigma$ redshift error
$\sigma_z/(1 + z)$ is $\sim 0.07$.  The comparison of the photometric redshift 
estimate and spectroscopic measurements shows, however, an increasing scatter from 
the one-to-one relation above $z_{phot}>0.8$ (see Fig. 4 in Wolf et al. 2004). 
This observed discrepancy between the photometric and the spectroscopic data 
at high $z$ is mostly due to the fact that COMBO-17 is based on optical SEDs only, 
and cannot exploit the Balmer-jump spectral feature for galaxies at $z>1$.
Altogether, of our 783 galaxies without spectroscopic redshift, we have used 
375 photometric estimates from COMBO-17, all those with $z_{phot}\leq 0.8$.
One hundred and 26 IRAC 3.6 $\mu$m sources do not have a COMBO-17 association
due to the different selection functions.

For the remaining 408 sources (27\%) in our sample, photometric redshifts 
have been estimated here using the $Hyperz$ code (Bolzonella et al. 2000), 
exploiting the availability of an extensive photometric coverage
at $\lambda> 1\mu$m. 
In particular we have included in our analysis the IRAC 3.6 $\mu$m fluxes, 
as providing a useful contraint on the solutions (e.g. 
Rowan-Robinson et al. 2005; Polletta et al. in prep).
While fine-tuning the procedure, it turned out that fairly accurate photometric
redshifts were obtained by using only two templates from the set provided by $Hyperz$, 
namely one for ellipticals ($E.ised$) and one for normal Sb spirals ($Sb.ised$). 
We assumed the reddening law by Calzetti (2000).

We have used the numerous spectroscopic redshifts in the field to check 
the reliability of our photometric estimates.   
For the spheroid (late-type) classes we obtained a median offset of
$\Delta(z)/(1 + z_{spec})$ = 0.06 (-0.08) and a rms
scatter of  $\sigma [\Delta(z)/(1 + z_{spec})]$ = 0.08 (0.11).   
In Figure \ref{zphot} we report a comparison of photometric versus spectroscopic redshifts 
separately for early-type (left panel) and late-type galaxies (right panel). We show
here sources with $S_{3.6}>$10 $\mu$Jy and for which $Hyperz$ found a reliable
fit (corresponding to a rejection probability based on $\chi^2$ lower than 99\%).
These plots show that our procedure implies some systematic deviation 
at $z<0.6$, where $Hyperz$ tends to underpredict the redshift.

For only 60 sources in the complete $10\ \mu$Jy sample either COMBO-17 or the 
$Hyperz$ fits turned out to be marginally acceptable or bad (rejection $\chi^2$ 
probability higher than 95\%). This however corresponds to only $\sim 4\%$ of 
the Spitzer complete sample, hence does not impact on our conclusions.

\section{STATISTICAL ANALYSES }\label{stati}

\subsection{Extragalactic Number Counts from Spitzer IRAC Data }
\label{counts_obs}

We have first estimated the extragalactic number counts at 3.6 $\mu$m
in the GOODS-CDFS area, by weighting each sources with the reciprocal
of the effective area $1/A_{\rm eff}(>S_i)$ 
within which they could be detected to a given flux density.
The errors associated with the counts in each flux level have been computed
as in Gruppioni et al. (2002):
\begin{equation}
\sigma_N = \sqrt{\sum_i{1/A^2_{{\rm eff}}(>S_i)}},
\end{equation}
where the sum is over all the sources with flux density $S_i$. These errors 
represent the Poissonian term of the uncertainties, and have to be
considered as lower limits to the total errors.

\begin{figure}
\centerline{
\psfig{file=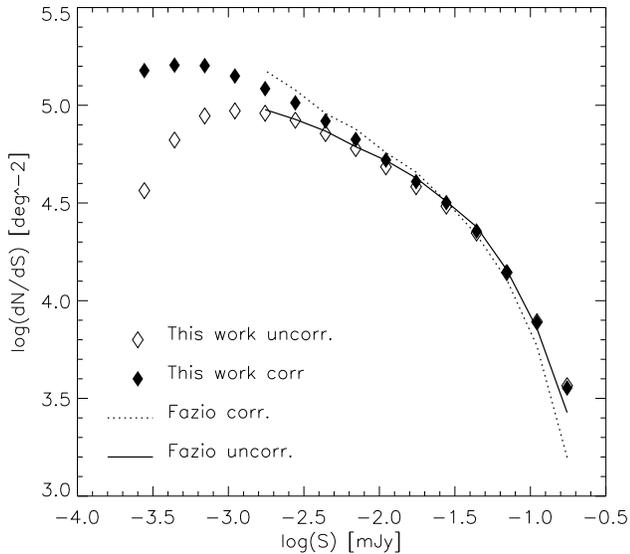,width=9cm}}
\caption{The IRAC/GOODS differential 3.6 $\mu$m counts from
this work are compared with those published by Fazio et
al. (2004). We report our incompleteness-corrected (filled diamonds)
and the uncorrected (open diamonds) data with the corresponding
corrected (dotted line) and uncorrected (solid line) counts 
by Fazio et al. (2004).}
\label{cfazio}
\end{figure}

We plot in Figure \ref{cfazio} a comparison of the IRAC/GOODS total differential 
3.6 $\mu$m counts from our analysis with those published by Fazio 
et al. (2004), that were computed from a much wider sky area ($\sim 9$ square
degrees). We report our counts corrected for incompleteness (filled diamonds)
and those uncorrected (open diamonds), versus the corresponding
corrected (dotted line) and uncorrected (solid line) counts 
by Fazio et al. (2004).

A general agreement is observed between the two
independent samples, in particular concerning the ``raw''
counts. Once the data are corrected for the corresponding selection
functions, we observe the Fazio's et al. distribution keeping slightly 
higher at the fainter flux range ($S_{3.6}<$3 $\mu$Jy). 
Below this level, source confusion starts to play some role, 
and it is difficult to assess if the reason for this slight 
discrepancy might be due different treatment of the blended sources
or differences in the determination of the completeness corrections.
In any case, this does have a quite modest impact on the 
integral counts which are used in the following.

\begin{figure*}
\centerline{
\psfig{file=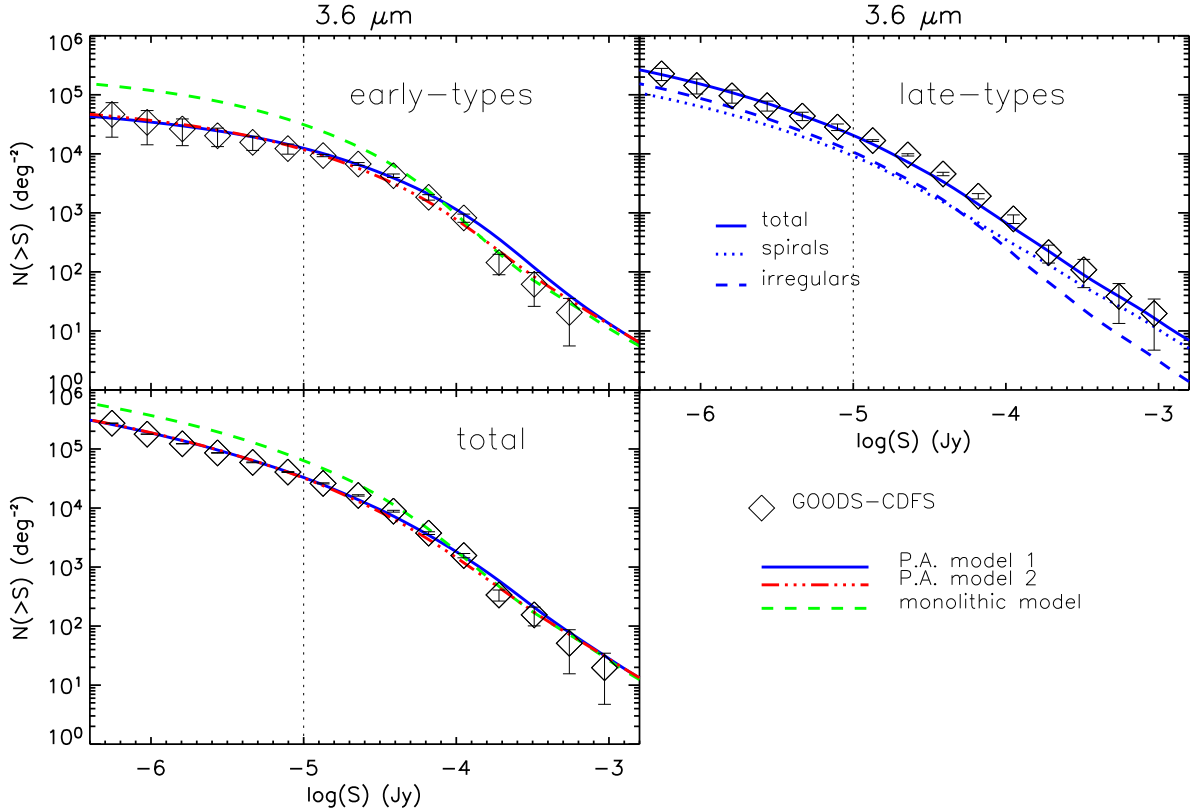,width=17cm}}
\caption{The 3.6 $\mu$m cumulative extragalactic number counts from the
IRAC/GOODS sample, corrected for incompleteness. The two upper panels 
show the relative contribution of the different morphological classes (early- and
late-types). The error bars include the contribution of the uncertainties
in the morphological classification (see Sect. \ref{morph}), which explains
their increasing size at the fainter fluxes.
The lower panel reports the results for the whole 3.6 $\mu$m band
population. The vertical dotted lines mark the 10 $\mu$Jy
limit, where the sample is $\sim 95\%$ complete, but the completeness
correction for the number counts is well controlled down to $\sim 1\ \mu$Jy
(see Fig. \ref{dr36}). The
data are compared with the predictions of our $Protracted$-$Assembly$ model 1
for spheroidal galaxies (solid blue lines), the $Protracted$-$Assembly$ model
2 (three dots-dashed red line) and the $monolithic$ model (dashed green lines).}
\label{cint36}
\end{figure*}

Figure \ref{cint36} provides details about the cumulative number counts 
for our IRAC/GOODS sample of 3.6 $\mu$m selected galaxies (empty diamonds) for
different morphological classes (all stars are excluded). 
The contributions of each
sources to both the counts and the associated errors are weighted for
the area within which the source is detectable. 

The two upper panels in the figure show the relative
contribution of the early- and late-type morphological classes,
while the lower panel the results for the whole 3.6 $\mu$m population. 
The errorbars include both the Poisson noise and the uncertainties
in the morphological classification as discussed in detail in Sect. \ref{morph}.
The former dominate at the brighter fluxes, while the latter
uncertainty determines the errorbars at the faint flux limits.
All information on number counts is also reported in Table \ref{tab:counts}.

\begin{table*}
\begin{center}
\begin{tabular}{c c c c c c }
\hline
$\log S_{3.6}$ & $n(>S_{3.6})$ & corr. fact. & $N_{TOT}(>S_{3.6})$ & $N_{E/S0}(>S_{3.6})$ & $N_{Sp/Ir}(>S_{3.6})$ \\
          (Jy) &               &             & ($sq. degree^{-1}$) & ($sq. degree^{-1}$)  & ($sq. degree^{-1}$)   \\

\hline
 -6.25& 5955& 2.05 &249731$\pm1508$ &  22608$\pm 27608$ & 227123$\pm 52786$ \\
 -6.02& 5095& 1.59 &165716$\pm1365$ &  21456$\pm 19930$ & 144260$\pm 40959$ \\
 -5.79& 4160& 1.36 &115744$\pm1206$ &  19398$\pm 12638$ & 96346$\pm 24134$ \\
 -5.56& 3343& 1.20 &82058$\pm1047$ &  16457$\pm 6982$  & 65601$\pm 12565$ \\
 -5.33& 2431& 1.15 &57206$\pm 882$&  13577$\pm 4227$  & 43629$\pm 7059$ \\
 -5.10& 1757& 1.10 &39551$\pm 727$&  11108$\pm 2390$  & 28443$\pm 3639$ \\
 -4.87& 1195& 1.07 &26158$\pm567$ &  9359$\pm 423$    & 16799$\pm 567$ \\
 -4.64&  761& 1.05 &16351$\pm 433 $&   6739$\pm 363$    & 9612$\pm 433$  \\
 -4.41&  419& 1.03 &8834$\pm 302$&  4269$\pm 292$    & 4565$\pm 302$   \\
 -4.18&  183& 1.00 &3758$\pm 199$&  1838$\pm 195$    & 1920$\pm  199$  \\
 -3.95&   78& 1.00 &1614$\pm 131$&   822$\pm 130$    & 792$\pm 131$   \\
 -3.72&   17& 1.00 &356$\pm 71$&   144$\pm  54$    & 212$\pm 71$   \\
 -3.49&    8& 1.00 &169$\pm 54$ &    61$\pm  35$    & 108$\pm 54$   \\
 -3.25&    3& 1.00 &58$\pm 35$&    20$\pm  15$    & 38$\pm 25$   \\
 -3.02&    1& 1.00 &19$\pm 15$&     0$\pm   0$    & 19$\pm 15$ \\
\hline
\end{tabular}
\end{center}
\caption{Galaxy number counts at 3.6 $\mu$m. Meaning of the columns: logarithm of the
limiting 
flux density in Jy; number of galaxies in the survey to the flux limit; incompletess
correction factor; integral number counts in sq. degrees$^{-1}$; number counts
for spheroidal galaxies; number counts for late-type galaxies (spiral, mergers, irregulars).
Uncertainties in the number counts include those from the morphological classification. }
\label{tab:counts}
\end{table*}

The vertical dotted lines mark the limit, 10 $\mu$Jy,
above which the sample is $\sim 95\%$ complete.
The data and their errorbars have all been corrected for incompleteness 
(see Section \ref{compl}).   The correction factor at the 1 $\mu$Jy limit
amounts to a factor 1.5 only (see Fig. \ref{dr36}).
The implications of the total observed counts and the separate contributions 
by the various morphological classes will be discussed in Section \ref{model}.

\subsection{Redshift Distributions }
\label{dz_obs}

\begin{figure*}
\centerline{
\psfig{file=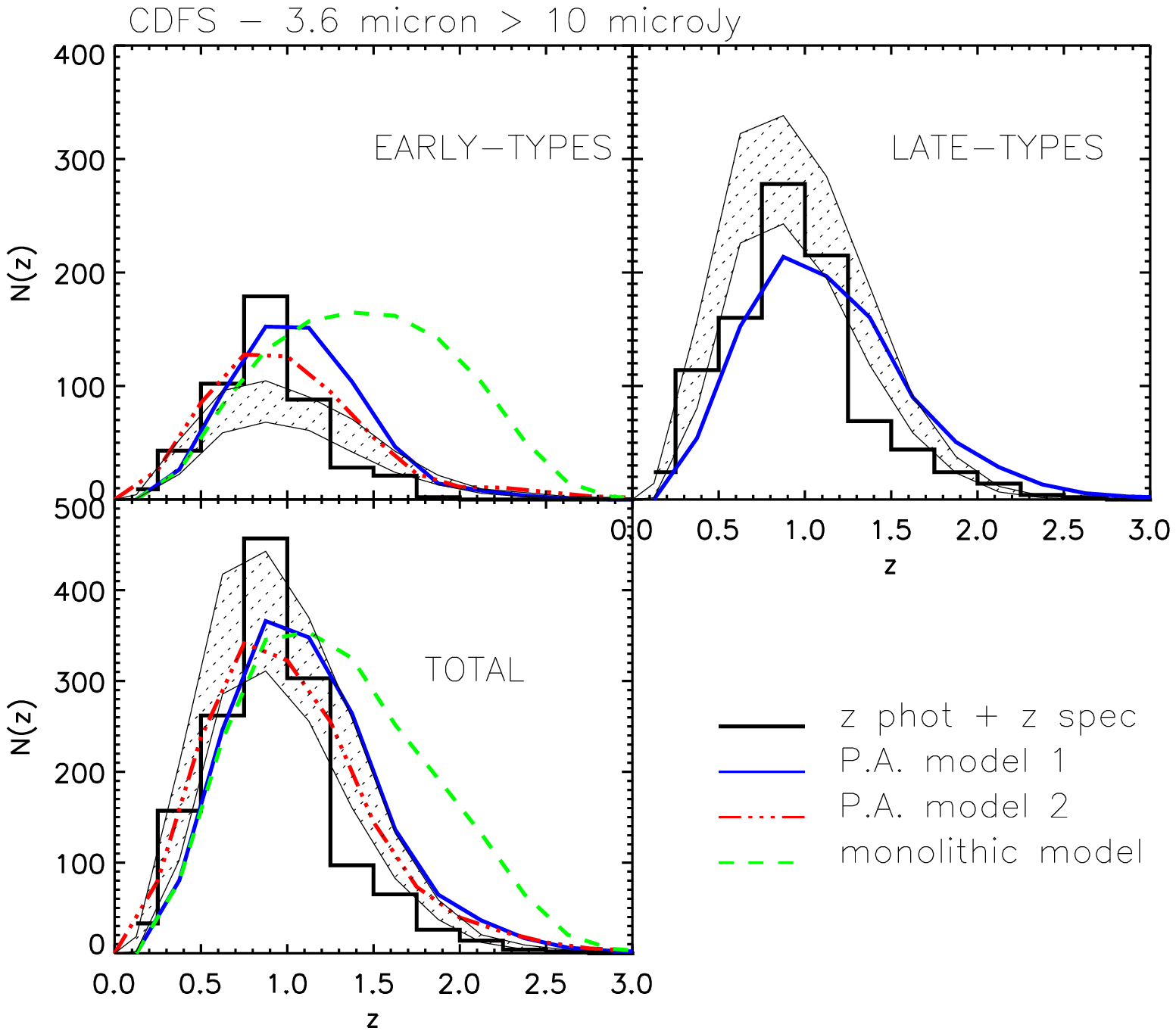,width=17cm}}
\caption{Redshift distributions of the 3.6 $\mu$m IRAC/GOODS
sample with $S_{3.6}>$10 $\mu$Jy over an area of 160 sq. arcmin, compared with 
model predictions. 
 Fourtyseven percent of the redshifts come from spectroscopic
observations (Cimatti et al. 2002, Le Fevre et al. 2004a, 
Vanzella et al. 2005, and this paper).
The upper panels show the relative contribution of spheroid and late-type 
galaxies, the lower panel reports the total distribution. 
The data are compared with the predictions of our $Protracted$-$Assembly$ 
model 1 (solid blue lines), $Protracted$-$Assembly$ model 2 (three dots-dashed 
red line) and our $monolithic$ model (dashed green lines). The hatched grey
regions correspond to predictions from the GALICS semi-analytical code,
illustrating the effects of cosmic variance over the sampled area. }
\label{Dz36}
\end{figure*}

The observed distributions of redshifts for complete galaxy samples 
provide a fundamental test for evolutionary models (e.g. Franceschini
et al. 1998; Somerville et al. 2004; Kitzbichler and White 2004).
      Figure \ref{Dz36} shows our estimated redshift distributions $N(z)$ 
for the 3.6 $\mu$m sample with $S_{3.6}>10\ \mu$Jy (thick solid 
histograms) in the CDFS area, a flux level at which 
the sample is 47\% spectroscopically complete. 
The distributions here include 695 spectroscopic, 375 COMBO-17 and 408 
$Hyperz$ photometric redshifts.       
The upper panels show the relative contribution of spheroids and
late-type galaxies, while the bottom panel reports the total distribution.
The prominent peak around $z\sim 0.8$ is  partly contributed by already
known CDFS galaxy overdensities falling within the redshift bin
(Cimatti et al. 2002, Vanzella et al. 2005, Le Fevre et al. 2004a, Adami
et al. 2005). 
The total observed distribution shows a rapid decline above $z\sim 1.2$, 
and this feature is common to both morphological classes. 
Elliptical galaxies are virtually absent above $z\sim 1.6$, while 
spiral/irregular galaxies show an apparent tail extending up to $z\sim 2.7$.

 Note that the detailed behaviour of the $N(z)$, as well as of the 
number counts, is a combination of the 
intrinsic evolution of the luminosity functions, the
K-corrections and the flux limit (e.g. Ilbert et al. 2004). We will
resort to detailed modellistic fits in Sect. \ref{model} to disentangle 
these various effects. 

Particular interest resides in the high-redshift tail of the 
distribution, because the statistics of the higher-z galaxies 
provides the tightest constraints on the formation models. 

A question might arise here about the reliability of the 
morphological classification for such faint distant objects.
We have then spent some effort in adding visual inspection to the automatic 
analyses for all 183 galaxies with $z>1.3$ in the complete sample.
This inspection has essentially confirmed the results of the 
CAS classification and has shown that 5-6\% at most of the 151 late-types 
could be misclassified spheroids, while $\sim$30\% of the 32 spheroids could 
be instead be classified as later type galaxies. Then, at close inspection,
we do not expect more spheroidal galaxies at $z>1.3$ than shown in Fig. \ref{Dz36}.

%
%
%
%
%
%

Our total observed $z$-distribution is consistent with the results of the analysis 
of Somerville et al. (2004) based on the photometric redshifts by 
Mobasher et al. (2004).
The Somerville's et al. analysis considered the z-distributions for only the total
population of a K-band selected galaxy sample.

We find also fair consistency with a recent report by Rowan-Robinson et al. 
(2005) based on the analysis of SWIRE survey data (Lonsdale et al. 2004) 
in the Lockman verification field (VF) to $S_{3.6}>$10$\mu$Jy and
$r < 25$, based on purely photometric redshifts and galaxy 
classification based on colors. Here, elliptical galaxies 
(red sub-population) also appear to cutoff at $z\sim 1.4$, while a 
tail of star-forming galaxies is seen up to $z\sim 3$. 

Our advantage compared to these analyses is in the substantial 
spectroscopic redshift coverage and the detailed structural analysis
of the Spitzer galaxy sample made possible by the very deep multiband ACS data.
A comparison with model predictions will be discussed in Sect. \ref{model}.

\begin{figure*}
\centerline{
\psfig{file=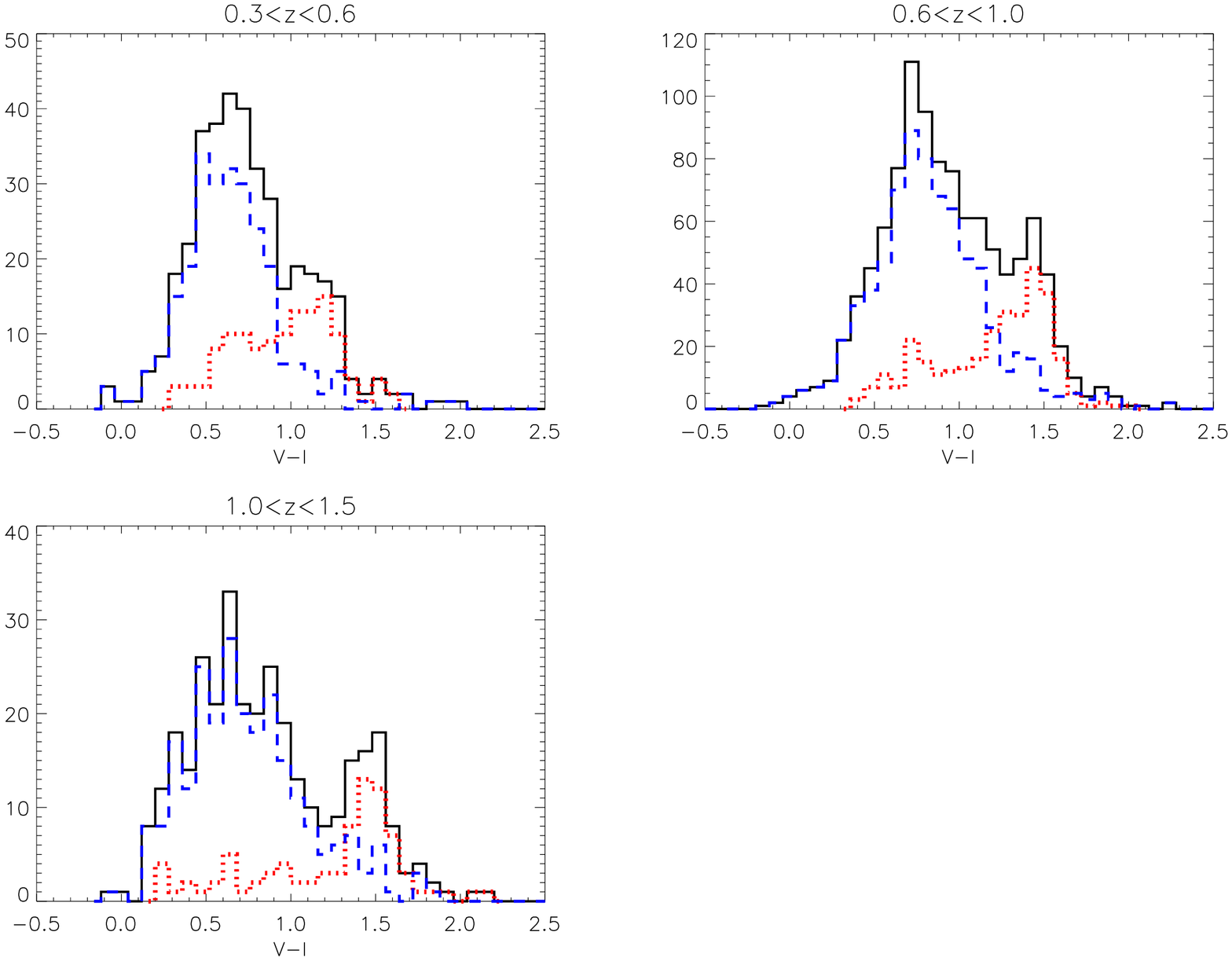,width=12cm,angle=90}}
\caption{Observed color bimodality of IRAC-selected galaxies 
in three different redshift bins. The red dotted line and the blue
dashed line correspond to our morphologically classified elliptical and
late-type galaxies, respectively, showing a clear correlation 
between the early-type systems and the reddest ones.
A substantial fraction of spheroidal galaxies, however, display
the colors of bluer spirals/irregulars, rather independent of redshift.}
\label{bimodality}
\end{figure*}

\subsection{Color Bimodality versus Morphological Classification }
\label{bimod}

Recent work has shown that the rest-frame color distribution of galaxies is
bimodal at all redshifts out to $z\sim 1$ (see, e.g. Hogg et al. 2002;
Blanton et al. 2003; Bell et al., 2004). One peak is red and consists 
of mostly quiescent galaxies earlier than Sa in morphological
type. The other peak is blue and consists primarily of star-forming
galaxies later than Sa (Strateva et al. 2001, Bell et al. 2004).

With the present large dataset we are able to check if our morphological
and structural classification is consistent with a color-based selection.
Figure \ref{bimodality} shows our observed color distributions from the 
$S_{3.6}>10\ \mu$Jy flux-limited sample in three
redshift bins from $z=0.3$ up to $z=1.5$. 
Remarkably, a best color differentiation turned out to be provided by the
V-I (or alternatively V-z) observed color in all redshift bins.
The red dotted line histograms correspond to the morphologically 
classified ellipticals, while the blue dashed line to the
late-type galaxies. 

These plots show not only a bimodal distribution at all redshifts,
but also a clear correlation between galaxies with early-type 
(late-type) morphological classification and the reddest (bluest) 
population.
There is only a marginal trend for the early-type population to 
increase the median color with $z$, while the late-types keep 
remarkably unchanged.

However, it is relevant to note that quite a significant fraction, 
of the order of $\sim30\%$, of the population morphologically classified 
as spheroidal galaxies (472 objects brighter than 10 $\mu Jy$) 
display blue colors typical of later morphological 
types. Also, this result looks largely independent of redshift.

\begin{figure*}
\centerline{
\psfig{file=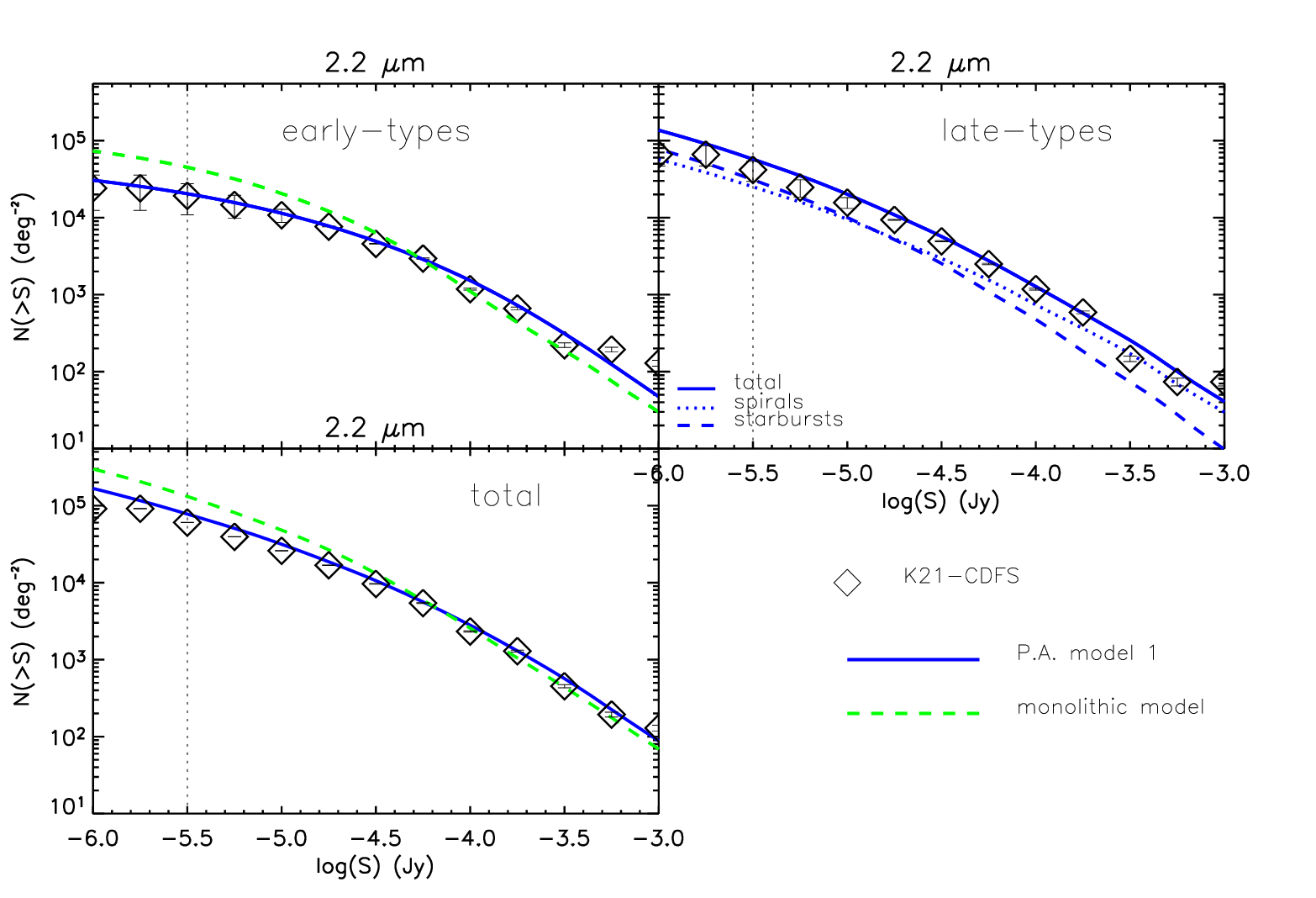,width=17cm}}
\caption{2.2 $\mu$m cumulative extragalactic number counts from the
IRAC/GOODS sample. The two upper panels show the relative contribution
of the two different morphological classes (early- and
late-types). The lower panel reports the results for the whole K band
population. The vertical dotted lines mark the K=21 mag (Vega
system) limit, where the sample is more than 90\% complete. The data
are compared with the predictions of our $PA$ model 1
(solid red lines).}
\label{cintK}
\end{figure*}

\begin{figure*}
\psfig{file=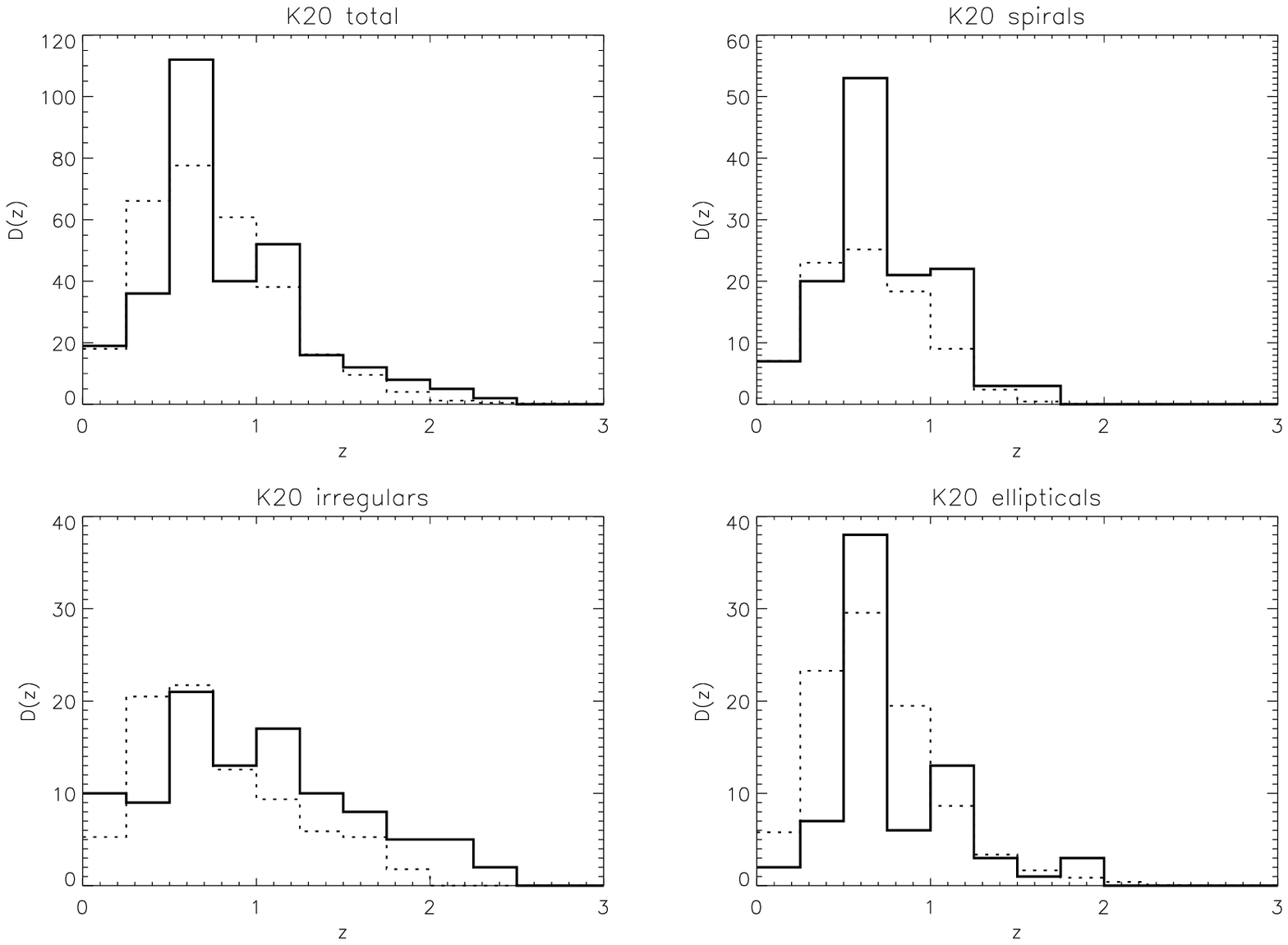,width=16cm}
\caption{Redshift distributions from the K20 catalogue, including 300 
morphologically classified galaxies over an
area of 32 sq. arcmin in CDFS (see CA05).
The solid line histograms in the four panels 
show the relative contributions of three morphological types as 
classified in CA05, the bottom left panel shows 
the total distribution. 
The dotted line histograms mark the prediction for our best-fit 
evolutionary model, in particular the $PA$ model 1
for spheroidal galaxies.}
\label{dzK20}
\end{figure*}

We have looked in some detail into this population of spheroidal
galaxies with blue colors, in particular considering the $\sim$150 such objects 
with $V-I<1$. Of these, roughly a third appear as typically bulge-dominated
early-type spirals or irregulars with bright compact cores, which are 
misclassified as spheroidal galaxies.
Roughly another third are normal ellipticals with blue colors, while the
remaining third are ellipticals with signs of interactions. 
Less than 1\% are compact objects.


Our conclusion is that a structural analysis of faint galaxies
provides independent information and classification tool
with respect to the most usually adopted multi-color data.

\subsection{Further Statistical Constraints from Deep K-band 
Observations}
\label{isaac}

We take advantage of the deep K-band data in the GOODS/CDFS field
to compare the statistical properties of the 
Spitzer long-wavelength galaxy population with those selected
at 2.2 $\mu$m (see Sect. \ref{Kimag}).
Figure \ref{cintK} shows the cumulative number counts
for our ISAAC/CDFS 2.2 $\mu$m selected galaxy sample
(thin black histogram).
Stars are excluded from the analysis. 
The two upper panels show the relative contribution of the different 
morphological classes (including uncertainties in the morphological
classification). The morphological analysis is the same 
as performed for the 3.6 $\mu$m sample, and the classification
criteria are consistent with those in CA05 (see Sect.\ref{morph}). 

The bottom panel in Fig.\ref{cintK} reports the results for the whole K band
population. The vertical dotted lines indicate the K=21 mag (Vega
system) limit, where the sample is more than 90\% complete.

We also derived redshift distributions for the $K$-band samples, 
taking advantage by the overlap between GOODS-CDFS, the VVDS (Le Fevre 
et al. 2004a) and K20 (Cimatti et al. 2002a; Mignoli et al. 2005) 
surveys, and by the fact that these are characterized by high spectroscopic 
completeness. 

The results are summarized in Figure \ref{dzK20}, where
the solid histograms in the four panels 
show the relative contribution of the three morphological types as 
classified by CA05 (including ellipticals, spirals, and irregulars/mergers,
which are differentiated here from spirals).
Model predictions will be compared with these data in Section \ref{model}.

\subsection{The Space-Time Distribution of the Sample Galaxies}
\label{vvmax}

We have further investigated the distribution in space-time of our sample 
galaxies by applying the $V/V_{max}$ test, first developed by Schmidt (1968)
to study the spatial uniformity of quasars.  The test compares the maximum 
comoving volume $V_{max}$, within which each sample objects should be visible, 
with the volume $V$ occupied by the source to the observed redshift.
For a uniform source space distribution, the $V/V_{max}$ values are
uniformly distribuited between 0 and 1, and $\left<V/V_{max}\right>=0.5$ within the
statistical error given by $1/(\sqrt{12\cdot N})$, $N$ being the number
of objects.

The volume $V$ within the survey area $\Omega$ and the redshift $z_1$
is computed as 
$$ V(z_1)=\Omega/(4\pi) \int_0^{z_1} \frac{dV}{dz}dz, $$
where $\frac{dV}{dz}dz$ is the differential comoving volume element.
The volume $V_{max}\equiv V(z_{max})$ refers to the maximum redshift $z_{max}$
at which the source would still be detected with a flux density matching
the survey limit. 
To compute it, we have first obtained a best-fit spectral representation 
($L_\nu$) of the observed SED,
as discussed in Sect. \ref{sfit} below, and the best-fit 
rest-frame monochromatic luminosity $L_{3.6}$ (or $L_K$ in the case
of the K20 sample discussed in Sect.\ref{vvk20}) at the source redshift $z$. 
Then the redshift of the source is increased from the observed value $z$ 
to that at which the flux density coincides with the survey
limit, using the luminosity-distance relation:
\begin{equation}\label{dL}
S_{\nu} = {L_{\nu} \over 4\pi d_L^2 K(\nu) }
\end{equation}
$d_L(z)$ being the luminosity-distance at redshift $z$ and $K_\nu$ the K-correction:
\begin{equation}\label{K}
K_{\nu} = (1+z) {\int L_{\nu(1+z)} T_\nu d\nu \over \int L_{\nu} T_\nu d\nu } .
\end{equation}
In eq.(\ref{K}) the best-fit spectral representation $L_\nu$ is kept
fixed as a function of redshift, without applying any evolutionary corrections.
In the case that the source luminosity $L_\nu$ would increase at higher-z 
due to the younger ages of the stellar populations (which might be 
considered as typical), the volume $V_{max}$ available to the source would be 
underestimated (which then provides us with an upper limit to $V/V_{max}$).

\begin{figure*}
\begin{center}
\psfig{file=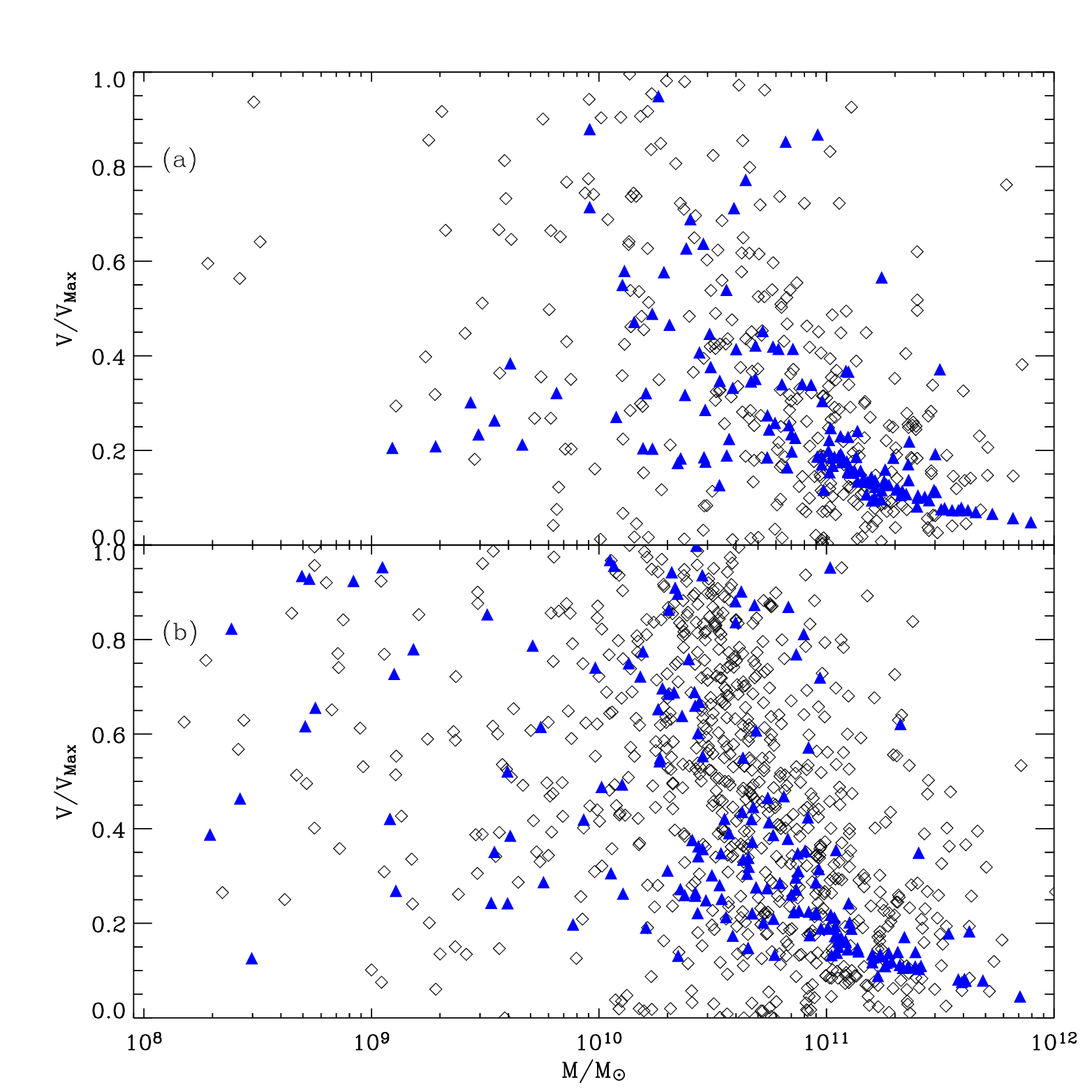,height=15cm,width=17cm}
\end{center}
\caption{
$V/V_{max}$ values for the early-type ({\sl panel a}) and late-type ({\sl panel b}) 
galaxies in the sample as a function of the stellar mass (estimated as discussed in
Sect. \ref{sfit}) based on the flux-limited $S_{3.6}>10\ \mu Jy$ IRAC sample. 
Filled symbols indicate the objects in the vicinity of the clusters at z=0.647 and
z=0.735.  Mean and median values of $V/V_{max}$ for various subsamples are reported
in Table \ref{tab:vvmax}.
}
\label{vmax_goods}
\end{figure*}

\subsubsection{The $V/V_{max}$ test for the Spitzer 3.6 $\mu$m galaxy sample}

We have first applied the $V/V_{max}$ test to the objects in the flux-limited
GOODS/IRAC sample with $S_{3.6}\geq 10\ \mu$Jy. 
   In spite of the limited spectroscopic coverage (47\%) for the 
Spitzer-selected sample,
we believe that we can still obtain a significant assessment of the source space 
distribution:
the use of the mean of the $V/V_{max}$ statistics is expected to average out 
random errors (though not the systematic ones) introduced by the photometric 
redshift estimate.  

Due to the moderate angular resolution ($\sim 3\ arcsec$) of the Spitzer IRAC 3.6 $\mu$m 
images, the high-redshift detected galaxies are spatially un-resolved and can 
be considered as behaving like point sources. For these, the 
effects of the cosmological angular size variations and surface brightness
dimming ($\mu \propto [1+z]^{-4}$) do not directly affect the source detectability.
For this reason, we have not attempted to carry out detailed simulations 
of all selection effects, not useful in the present situation, and rather
computed $z_{max}$ by simply considering the cosmological dimming of the 
total flux as in eq.(\ref{dL}).

\begin{figure*}
\centerline{
\psfig{file=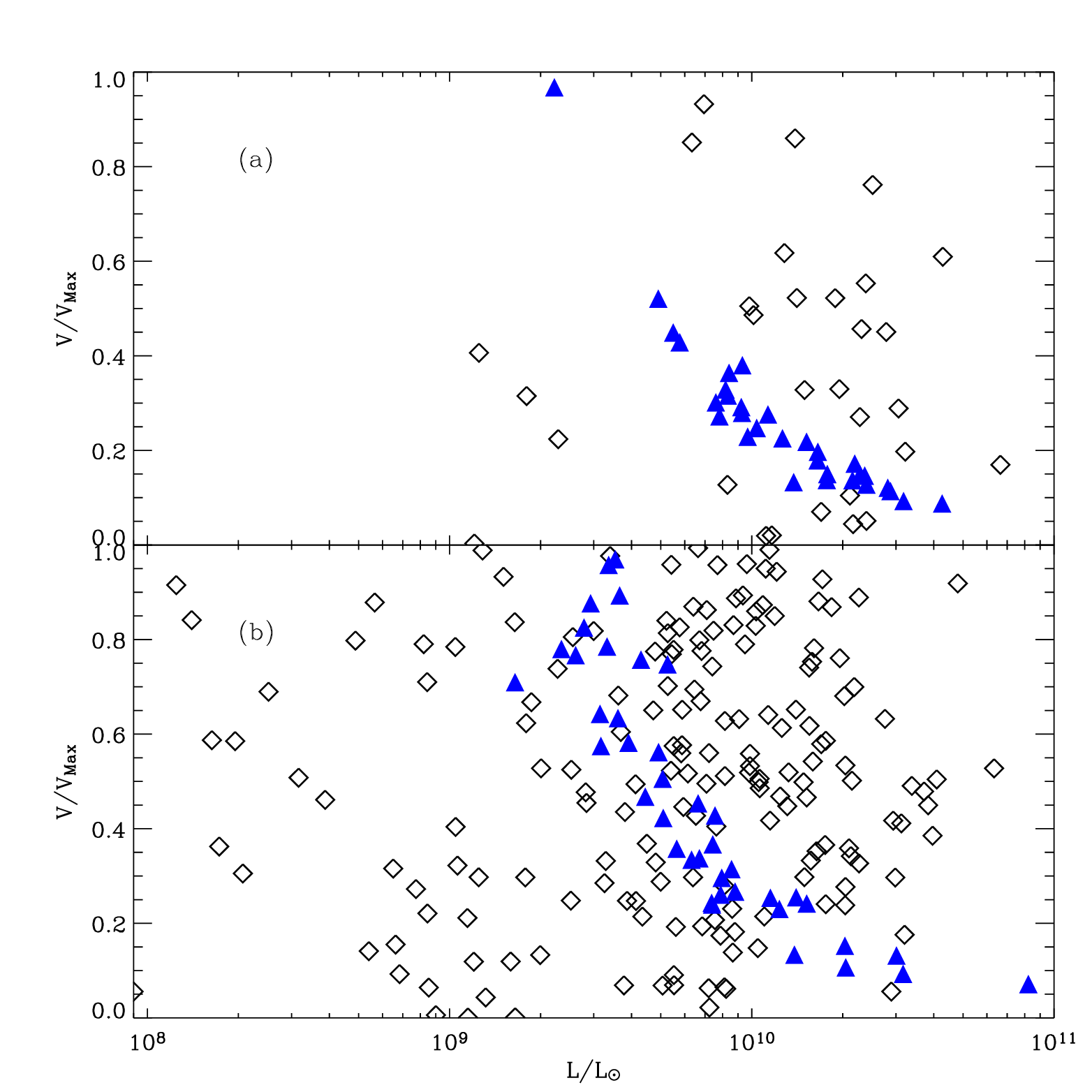,height=10cm,width=14cm}   }
\caption{
$V/V_{max}$ values for the early-type (panel a) and late-type (panel b) 
galaxies in the sample as a function of the K-band luminosity
from the K20 sample. 
Filled symbols indicate the objects belonging to the cluster at z=0.735. 
Mean and median values of $V/V_{max}$ for various subsamples are reported
in Table \ref{tab:vvmax}.
}
\label{vmax_k20}
\end{figure*}

A problem potentially affecting our analysis of the source space distribution 
are the obvious galaxy clusters/aggregations peaking at z$\sim$0.67 and 0.73 and 
visible in both the VVDS (Le Fevre et al. 2004a) and GOODS/IRAC (see Fig.
\ref{zmass} below) catalogues, 
which may bias our assessment of the sample homogeneity in redshift space.
For this reason, and in addition to the total flux-limited sample, 
we have considered a second one excluding sources (and the
corresponding $V$ and $V_{max}$ volumes) between $z=0.63$ and $z=0.77$.
We have {\sl a posteriori} verified that 
this is a large enough redshift interval to get rid of the dispersion effect 
in $z$ introduced by the fraction of sources with photometric redshifts.

Figure \ref{vmax_goods} reports our results on the distribution of the 
$V/V_{max}$ statistics for the spheroidal and the late-type
populations as a function of the galaxy's stellar mass 
(see Sect. \ref{sfit} below). 
The blue filled triangles correspond to sources falling 
in the $z$=0.63 to 0.77 redshift interval, which show a correlation
between the $V/V_{max}$ and the mass values (or with the luminosity in the subsequent
Fig. \ref{vmax_k20}). This correlation is due to the fact that the more massive and 
luminous objects at a given redshift would be visible over a wider $V_{max}$ volume.
A similar effect, with larger scatter, is also apparent in the whole
population, because of the characteristic peak in the source redshift 
distribution (Fig. \ref{Dz36}).

The mean and median values of $V/V_{max}$ for the various
galaxy subsamples are summarized in Table \ref{tab:vvmax}.
As shown there, the $V/V_{max}$ distributions for both the spheroidal 
and the star-forming classes show significant 
departures from the uniform distribution's expectation value. 
The most significant effect is the apparent dearth of spheroidal galaxies at 
high-z indicated by the very low $\left<V/V_{max}\right> \sim 0.32 \pm 0.03$, 
(median value of $\sim 0.25\pm 0.03$), obtained both including and
excluding the cluster volume. 
    Although formally this departure from uniformity is very significant,
we should caution however that the size of our field is relatively small 
and our $V/V_{max}$ analysis subject to cosmic variance effects,
particularly for the more strongly clustered spheroidal population.   

Late-type galaxies show less evident departure from uniformity. Within this 
broad galaxy ensemble, we have attempted to check if differences might be present
in the $\left<V/V_{max}\right>$ between early-type spirals and late-type 
spirals and irregulars.
We have used for this the asymmetry parameter $A$ and set the value of
0.4 as discriminating the most irregular systems from more symmetric
standard spiral galaxies (e.g. CA05; Conselice 2003b).
Table \ref{tab:vvmax} indeed reveals that regular spirals with $A<0.4$ 
tend to show a somewhat similar, though
lesse extreme, dearth of objects at the survey upper redshift boundary,
whereas merger/irregulars reveal a marginal evidence for positive
evolution ($\left<V/V_{max}\right>\simeq 0.513\pm 0.005$).
This result is in good agreement with the value of $\left<V/V_{max}\right>$
in excess of 0.5 found by Daddi et al. (2004b) for star-forming (starburst
and irregular) objects at $z>1.4$.

\begin{table*}
\begin{center}
\begin{tabular}{c c c c c}
\hline
\hline
Gal. population (GOODS)  & spheroidals   &  late-types    & late-type with $A>0.4$ & late-type with $A<0.4$ \\
\hline
Number of sources        & 465           &  949           & 316                    & 633                    \\
 cluster incl.           & 0.318 (0.252) &  0.452 (0.424) &   0.502 (0.496)        & 0.403 (0.353)          \\
 cluster excl.           & 0.320 (0.245) &  0.445 (0.416) &   0.513 (0.504)        & 0.408 (0.362)          \\
\hline\hline
Gal. population (K20)    & spheroidals   &  late-types    & late-type with $A>0.4$ & late-type with $A<0.4$ \\
\hline
Number of sources        & 61            &  230           & 37                     & 189                    \\
 cluster incl.           & 0.313 (0.272) &  0.499 (0.439) & 0.430 (0.385)          &   0.426 (0.332)   \\
 cluster excl.           & 0.370 (0.330) &  0.507 (0.464) & 0.519 (0.523)          &   0.444 (0.411)   \\
\hline
\end{tabular}
\end{center}
\caption{Average values of the $V/V_{max}$ test for various galaxy populations.
Median and mean values are reported inside and outside parentheses, respectively.  
The three upper and three lower
lines refer to the GOODS/IRAC and K20 galaxy samples, to the flux limits of $S_{3.6}=10\ \mu Jy$
and K=20, respectively. Statistics of $V/V_{max}$ are reported here for both the 
total GOODS/IRAC and K20 samples (cluster incl.) and after exclusion of
sources taking part in strong 
galaxy concentrations around z=0.7 cluster excl.).  The latter
for the GOODS/IRAC sample is achieved by excluding from the sample the
cosmic volume between $z=0.63$ and $z=0.77$.
}
\label{tab:vvmax}
\end{table*}

%

\subsubsection{$V/V_{max}$ for galaxies in the $K$-band selected sample}
\label{vvk20}

We have also applied the $V/V_{max}$ test to objects in the K20 sample 
($K_s\leq 20$) analyzed by CA05, including 74 E/S0 and 226 Spiral/Irregular
galaxies. Though having lower statistically significance, the advantage 
offered by this sample is the virtual completeness of its spectroscopic 
follow-up,  also useful in identifying and excluding
galaxies belonging to the z=0.7 cluster.   

Even in the $K_S$-band ISAAC images, with a typical spatial
resolution close to 1 arcsec, the vast majority 
of the sources do not appear as spatially resolved.
We have then calculated the $V/V_{max}$ distribution as for the
GOODS/IRAC sample, that is neglecting detailed treatment of
surface brightness dimming effects.
The results of the test, summarized in Figure \ref{vmax_k20} and 
Table \ref{tab:vvmax}, appear completely consistent with those from
the GOODS sample.

In conclusion,  we find significant evidence in our field that galaxies belonging 
to the normal Hubble sequence, 
i.e. spheroidals and spirals with low asymmetry,
tend to disappear towards the survey limit at $z\sim 1$ to 2,
while highly asymmetric objects (irregular/mergers) are more evenly
distributed.
Our results extend to low redshifts and strenghten the conclusion of Daddi 
et al. (2005a) that the $V/V_{max}$ distributions is skewed to low values
for the spheroidal population, indicative of negative evolution.
The characteristics of this evolution will be better specified 
and clarified in the following Sections.

\section{THE GALAXY EVOLUTIONARY MASS AND LUMINOSITY FUNCTIONS}
\label{mf}

The quality of data available in this area, particularly
the redshift information (either spectroscopic or photometric) and
the extensive photometric coverage of the galaxy SEDs, motivated us to
attempt a direct estimate of the evolutionary luminosity and mass 
functions for galaxies.

\subsection{The Broad-band Spectral Fitting Procedure}
\label{sfit}

\begin{figure*}
\centerline{
\psfig{file=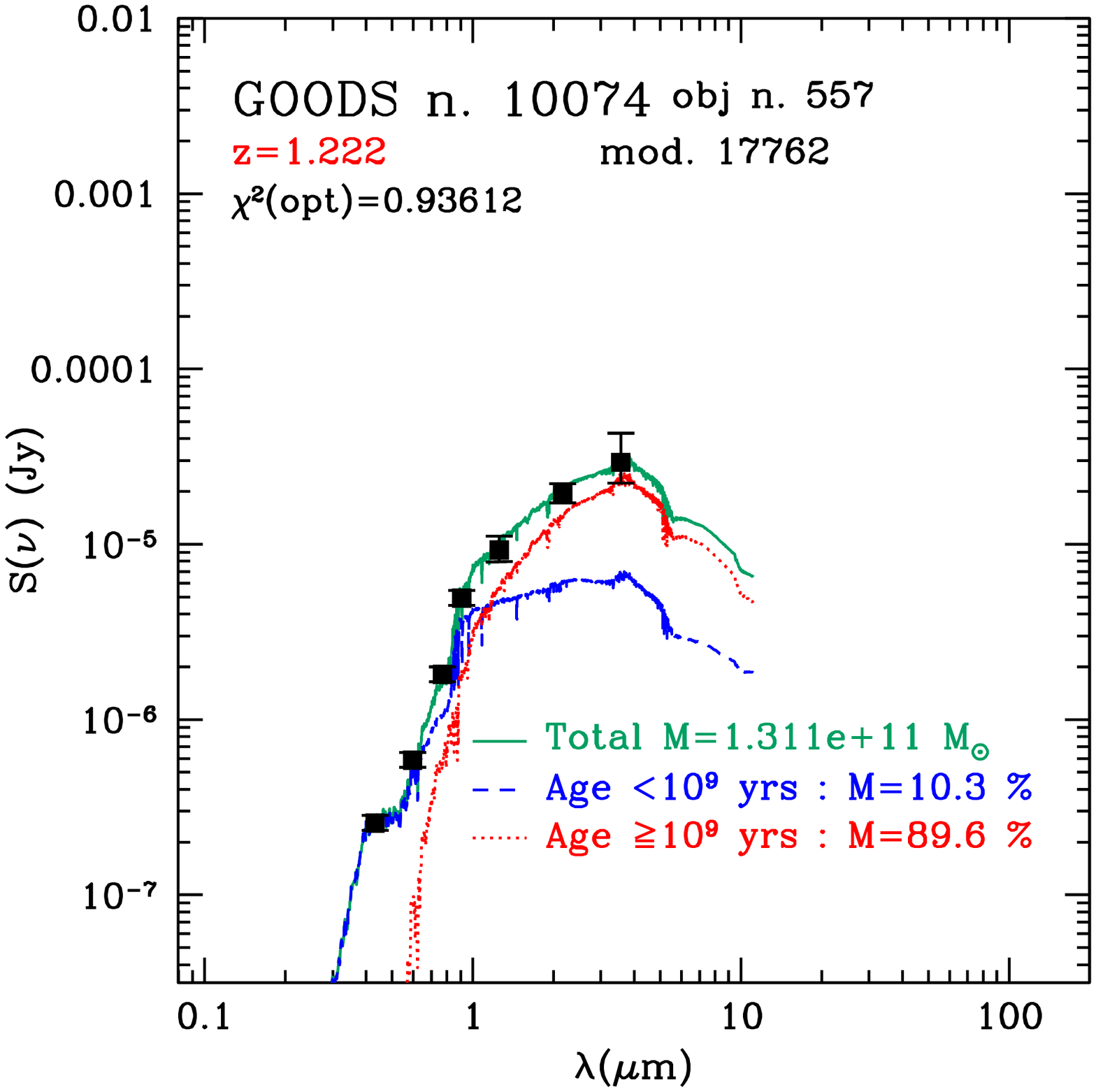,width=8cm} \hfil
\psfig{file=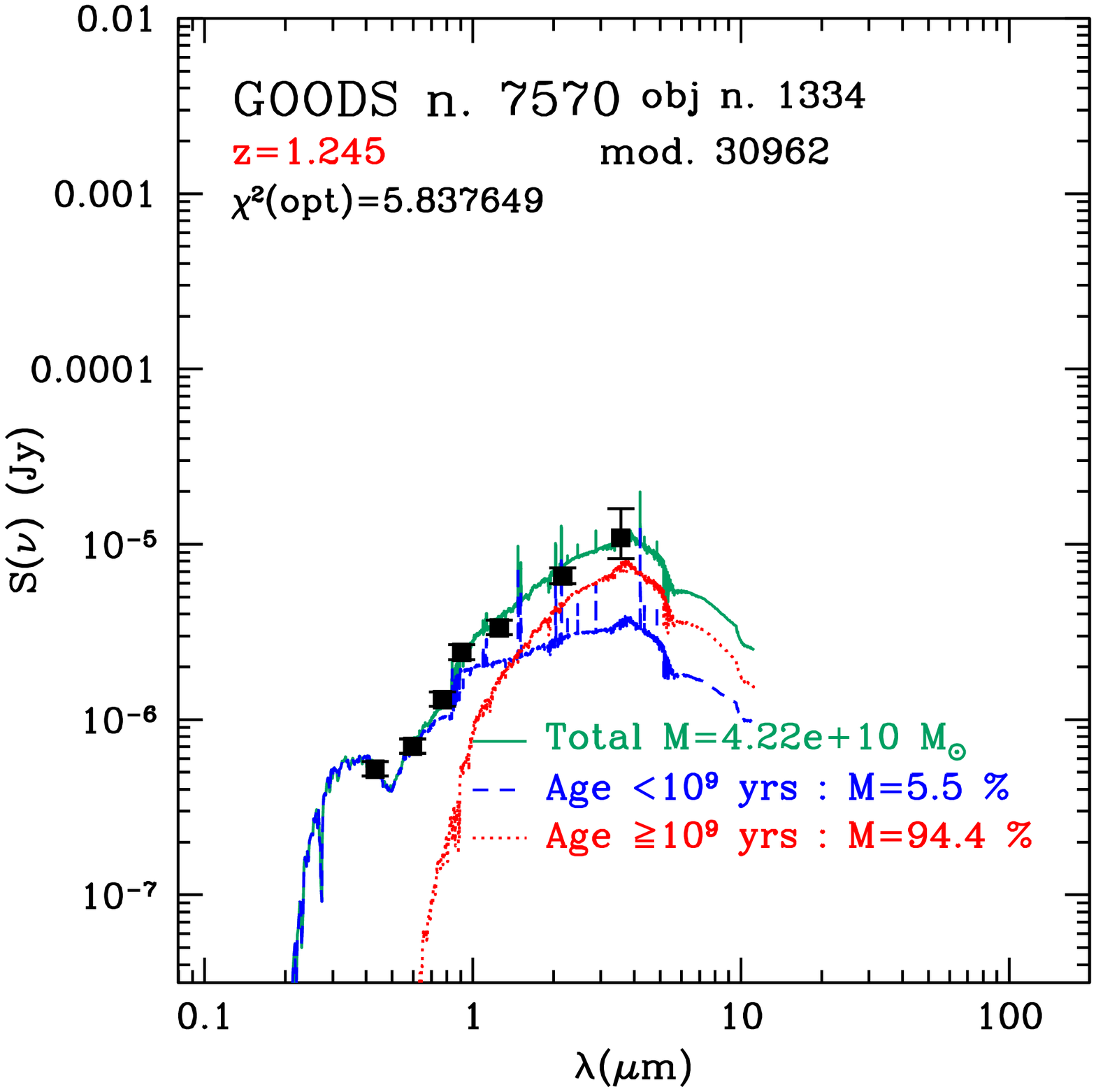,width=8cm}   }
\caption{ Examples of spectro-photometric fits to the observed SED of a
galaxy with elliptical  ({\sl left-hand panel}) and
and one with late-type morphology.  The main physical parameters are
reported in the figure labels, in particular the masses of stellar 
populations of different ages.
}
\label{examp}
\end{figure*}

Our estimate of the stellar mass follows from a detailed 
comparison of the observational SEDs with galaxy synthetic spectra. 
To generate them, we have adopted the spectrophotometric
synthesis code described by Berta et al. (2004), which is a development 
of that described in Poggianti, Bressan \& Franceschini (2001).
   We have made use of the latter, rather than relying on the stellar
mass estimated e.g. by the $Hyperz$ code (Bolzonella et al. 2000), 
for various reasons. The code by Berta et al. has been specifically
taylored to the stellar mass evaluation and has been systematically 
tested against the outputs of other codes (e.g. Dickinson et al. 2003;
Fontana et al. 2004). In addition, the code accounts in detail 
for the complex variety of stellar populations and population-dependent
extinction typical of star-forming and starburst galaxies, 
as detailed below.

According to Berta et al., 
the galaxy observed SED is modelled as a combination of a set of simple 
stellar populations (SSP) of solar metallicity and different age. 
Each SSP is weighted by its total mass and extinction, and represents
a temporal section in the star-formation history of the galaxy.
So, the mass fraction contained in any SSP corresponds to a given
averaged star formation rate (SFR) during the time section covered
by the SSP.
Each SSP is extinguished by a different amount of dust in a uniform 
screen and modelled
according to the standard extinction law ($R_V=A_V/E[B-V]=3.1$,
Cardelli et al. 1989). The total spectrum is built up by summing
the extinguished spectral energy distributions of all contributing
stellar generations.

Considering that high extinction values characterize only
stellar populations embedded in thick molecular clouds and that
disk populations are moderately absorbed ($A_V\le 0.5$, e.g. 
Kennicutt 1992; Kauffmann et al. 2003), we have limited the E(B--V) 
values for the populations with ages $\geq 10^9yrs$ to be less than 0.1.

All the SSP spectra have been computed with a Salpeter initial mass
function (IMF) between 0.15 and 120 $M_{\odot}$, adopting the Pickles
(1998) spectral library, extended and interpolated 
with the Kurucz's (1993) stellar atmosphere models,
following the work of Bressan et al. (1994) and Bertelli et al. (1994).
Photospheric stellar and nebular (line and continuum) emission has been
included through the ionization code CLOUDY (Ferland, 1996).
The result is corrected for cosmological dimming and for the
K-correction and compared to the observed SED by convolving with
the appropriate filter transmission curves.

The code seeks for the best-fit solution, by exploring the parameters
space through the Adaptive Simulated Annealing algorithm (Ingber et al. 
1989) and minimizing the difference between the observed data and the 
model measured by the $\chi^2$.
It takes of the order of 10 minutes CPU time on a PC to explore
the whole parameter space per galaxy.

\begin{figure*}
\centerline{
\psfig{file=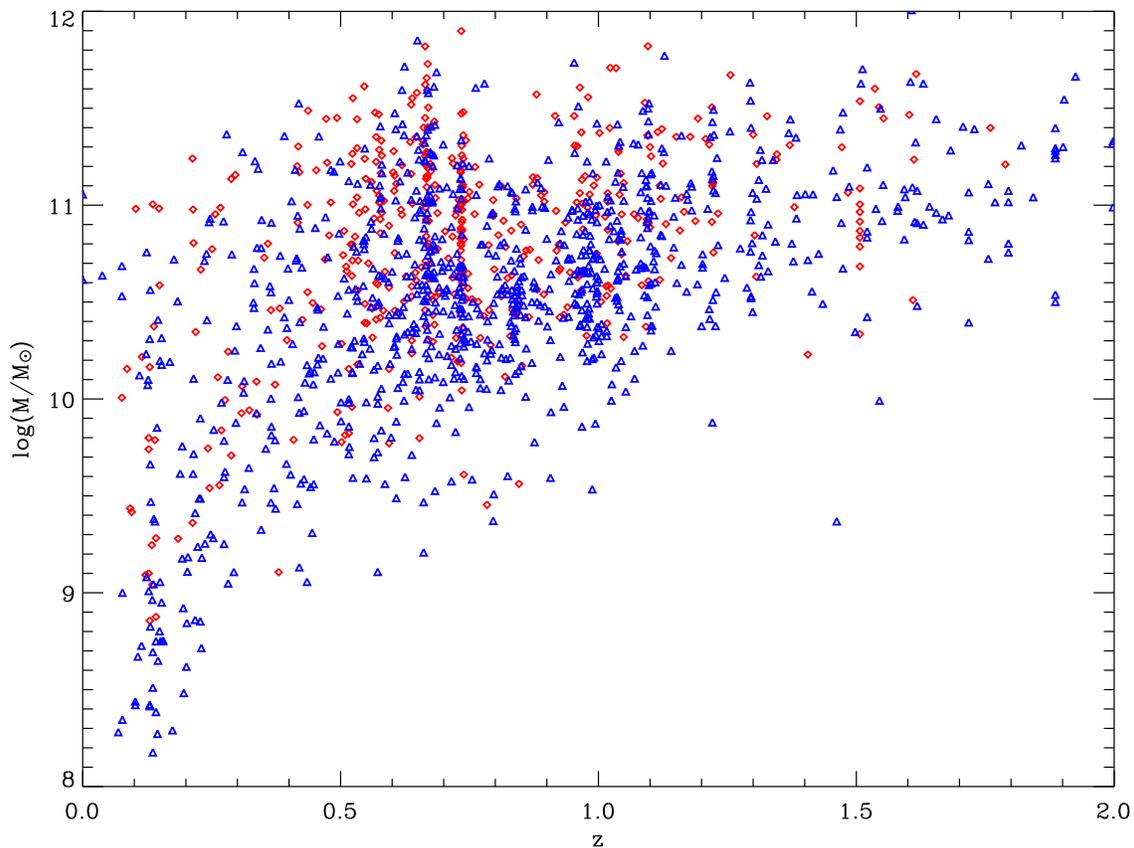,height=12cm,width=16cm}  }
\caption{Mass distribution as a function of redshift for the 3.6 $\mu$m IRAC/GOODS
  sample with $S_{3.6}>$10 $\mu$Jy. Morphologically classified
  ellipticals are marked as red diamonds, late-types as blue
  triangles. 47\% of the redshifts have a spectroscopic
  confirmation. Two structures at $z\sim 0.67$ and at $z\sim 0.73$ are
  particularly evident.}
\label{zmass}
\end{figure*}

The number of the SSP's involved in the fit depend on their
morphological classification and redshift. Ellipticals/S0 galaxies have
been fitted by using only SSP's older than $5\times10^8 yrs$. Late-type
galaxies have been modelled by adopting a combination of 9 SSP's, between
$1\times10^7 yrs$ and $12\times10^9 yrs$.
Obviously, for each galaxies only the SSP's younger than the age of the
universe at the galaxy redshift have been considered.

Our code outputs best-fit estimates for various physical parameters 
(e.g. rest-frame luminosity, age, star formation rate, extinction, 
stellar mass) for each sample galaxy. 
Due to the wide exploration of the parameter space,
we expect that not only the stellar mass value, but also the corresponding 
uncertainty should be fairly reliable, for a given stellar IMF.
This especially benefits by the Spitzer 3.6$\mu$m flux constraint
on the old stellar populations dominating the stellar mass,
particularly for the higher redshift ($z\geq 2$) galaxies.

The typical (2$\sigma$) uncertainties in the stellar mass determination
depend moderately on the source redshift (mostly thanks to the Spitzer
flux constraint) and are of the order of a factor 2 (somewhat less
for spheroidal galaxies in which the extinction is presumed to be
ineffective, somewhat more for star-forming in which the extinction 
adds to the total uncertainty).

We have applied our spectrophotometric fitting algorithm to the 
IRAC/GOODS 3.6 $\mu$m sample brighter than 10$\mu$Jy, using all 
the photometric bands available on
the multi-wavelength catalogue (see Sect.
\ref{multicat}), and the redshift. 
A couple of examples of the SED fitting quality are reported 
in Figure \ref{examp}, where the contributions by young and old stellar
populations are detailed.

In Figure \ref{zmass} we report the galactic stellar mass distribution 
as a function of redshift for the 3.6 $\mu$m IRAC/GOODS
sample with $S_{3.6}>10\mu$Jy. Morphologically classified
ellipticals are marked as red diamonds, late-types as blue
triangles. The two already mentioned structures at $z\sim 0.67$ and at 
$z\sim 0.73$, among others, are particularly evident.

\begin{figure*}
\centerline{
\psfig{file=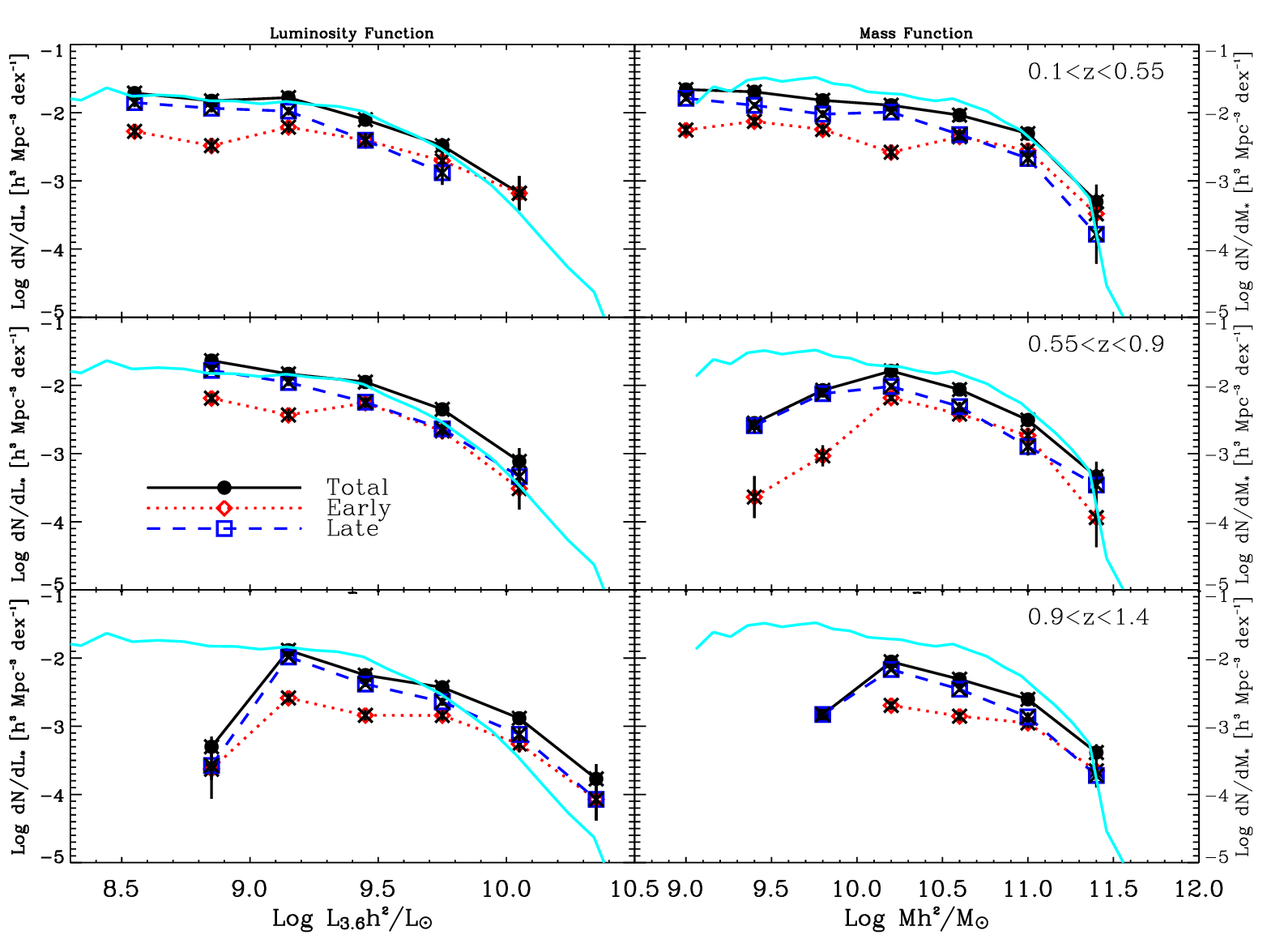,width=16cm,height=13cm}  }
\caption{Mass ({\sl right-hand panels}) and luminosity ({\sl left-hand panels}) 
function estimates derived from the 3.6 $\mu$m IRAC/GOODS
sample with $S_{3.6}>$10 $\mu$Jy, splitted into three redshift
bins from $z=0.1$ up to $z=1.4$. The contributions of the various
morphological classes is marked with different symbols:
early-types (open diamonds - dotted lines), late-types (open squares
- dashed lines), total (filled circles - solid lines). The thin solid 
line on the right marks the local mass function from Cole et al. (2001). 
In the intermediate redshift panels for both the mass and the luminosity
functions we have excluded all sources falling in the redshift interval
$0.63<z<0.77$, bracketing the cluster overdensities visible in 
Fig. \ref{zmass}. All plotted values of the mass and luminosity 
functions are expressed in terms of the $h=H_0/100\ Km/s/Mpc$ parameter.
}
\label{massfunc}
\end{figure*}

\subsection{Luminosity and Mass Functions}
\label{mfit}

Luminosity functions in various redshift intervals have been estimated
using the $1/V_{max}$ test, a standard method for flux limited
samples (e.g. Dickinson et al. 2003; Fontana et al. 2004; Bundy et al. 2005).
We have followed the same procedure to estimate $V_{max}$ as in Sect. \ref{vvmax},
except for the effect of the redshift binning: when $z_{low}< z < z_{high}$,
the $V_{max}$ is calculated as:
\begin{equation}
V_{max}=\Omega\int_{z_{low}}^{min(z_{high},z_{max})}\frac{dV}{dz}dz .
\end{equation}
Then the comoving number of galaxies for a given redshift bin a and for
a luminosity interval $\Delta L$ is estimated as:
\begin{equation}
\Phi(L)\Delta L=\sum_i\frac{1}{V_{max}^i}\Delta L .
\end{equation}\label{fi}

We report in Fig. \ref{massfunc} the luminosity functions 
($L_{3.6}=\nu_{3.6} L(\nu_{3.6}$) calculated for three 
redshift bins:  $0.2 < z < 0.55$, $0.55 < z < 0.8$ and $0.8 < z < 1.4$, containing
258, 471 and 503 galaxies, respectively. In each panels, the early- and late-type
contributions are plotted, together with the total functions (see the figure
caption for the meaning of the symbols). 
  Errorbars have been estimated by Poisson statistics. 
We have not attempted to include in our error budget the effects of 
uncertainties in the photometric redshift and stellar mass estimates, which
are overwhelmed by those related with the cosmic variance.
We excluded from the analysis redshift bins above $z=1.4$, since beyond this 
limit the spectroscopic completeness becomes low, the photometric redshifts are less 
reliable and the morphological classification more uncertain or even impossible. 
 Numerical values and errors for the luminosity functions are reported in
the Appendix.      

To take care of the two prominent structures at $z\sim0.67$ and 
$z\sim0.73$,  galaxies in the redshift interval  $z=0.63$ to 0.77
have been removed, as in Sect. \ref{vvmax}.

We then checked the level of agreement of our sampling 
compared to well established luminosity functions.
We report in each panels of Fig. \ref{massfunc}
the K-band local LF by Cole et al. (2001), transformed to 3.6 
$\mu$m using the rest-frame color of a typical sample galaxy and 
adapted for cosmology. This transformation is rather straightforward 
because the spectra of galaxies are all very similar to each other in this 
spectral range (we find $\left<S_{2.2}/S_{3.6}\right>\simeq 10^{0.24}$).
We see an excellent agreement between our LF in the lowest-redshift
bin and the Cole et al. local LF, except for a slight excess in the 
highest luminosity bin, likely due to evolution. 
The low-z function at the higher luminosities is dominated by early-type 
galaxies, in agreement with Bundy et al. (2005), Kauffmann et al. (2003),
and Croton et al. (2005).

The trend for increasing luminosity continues with increasing
redshift and gets quite significant in the z=1.2 bin. 
 At this epoch our most massive galaxies are $\sim0.7$ mag brighter 
on average than at $z=0$, in good agreement with the estimate by
Treu et al. (2005a) based on a detailed study of the evolution of
the early-type galaxy fundamental plane.
A likely interpretation for this increase in luminosity may be 
a decrease in the $M/L$ due to the dominant stellar populations in
galaxies getting younger with redshift.  
It is hard to establish from the left panels of Fig. \ref{massfunc} if 
there might be density evolution accompaning this evolution in $L$.

%
%
Note that the K-correction in the computation of the 3.6$\mu$m 
luminosity has an increasing effect at increasing redshift.  
However we believe that the corresponding uncertainties should not 
affect appreciably our LF determinations, since the near-IR galaxy spectra 
are rather well behaved and homogeneous in the relevant wavelength interval 
(see Fig. \ref{examp}).

The next step for obtaining the mass functions simply requires to exchange 
the $L$ with $M$ in eq. \ref{fi}, and binning in mass rather than in luminosity.
The results of this operation are reported in the right-hand panels
of the figure   and in the Appendix.    
The local mass function by Cole et al. (2001) is also reported for reference 
in all panels.

In spite of the different criteria for the sample selection, our mass 
functions are compatible with the results by Fontana et al. (2004) 
and Bundy et al. (2005), who found little evolution up to $z\sim1$. 
The Fontana et al. sample is $K$-band selected, has a better
spectroscopic redshift coverage ($\sim 95\%$), but is shallower
and over a smaller area (only 80 $arcmin^2$), while the Bundy et al.
sample is mostly optically selected and covers a larger area.

While the total mass function in the low-redshift bin at median $z\sim 0.3$ in 
Fig. \ref{massfunc} keeps marginally low with respect to the Cole et al. local 
estimate (which might indicate either a moderate evolution in comoving density
already at such low redshifts, or more likely a mismatch between Cole et al.
and our stellar mass estimates), we notice an extremely good match of the
low-$z$ luminosity functions. We take this as to support the reliability 
of our results, which is an important check on consideration of the 
uncertainties in the photometric redshift for a significant
fraction of our low-z sources and of the cosmic variance problem.

The evolutionary mass functions in Fig. \ref{massfunc} reveal some remarkable
differences compared to the luminosity functions. While the latter show essentially a
migration along the L-axis (luminosity evolution), the mass function is more
consistent with mere evolution in number density progressing from the lowest
to the highest $z$-bin.

\section{MODELLISTIC ANALYSIS }
\label{model}

We investigate in this section the effectiveness of number counts and redshift
distributions in further constraining the evolutionary properties of faint 
high-redshift galaxies.

\subsection{Simple Heuristic Models for Galaxy Counts and Related Statistics}
\label{ourmodel}

We attempt here to compare our statistical observables 
with simple heuristic prescriptions. 
In our approach, essentially three main galaxy classes are considered as 
dominating the near-IR selected galaxy catalogues and being characterized
by potentially different evolutionary histories: spheroidal (E/S0) galaxies, 
quiescent spirals, and an evolving population of irregular/merger
systems (hereafter the starburst population).

It is straightforward to show (e.g. based on the luminosity functions of
Seyfert galaxies and quasars extrapolated from the optical)
that active galactic nuclei do not significantly contribute to
the extragalactic counts in the IR (Franceschini et al. 2005; Bell et al.
2005b). So we have not considered further in our analysis 
the contribution of active nuclei.

The SED templates describing the spectral shapes at different
galactic ages, needed to calculate the K-corrections and to transform
the luminosity functions from one wavelength to the other, 
have been computed using the same spectral synthesis code 
based on the Padova stellar isocrones as in Berta et al. (2004).
Consistently with Sect. \ref{sfit},   
we have adopted a Salpeter IMF with a lower limit $M_l=0.15 M_{\odot}$
and a Schmidt-type law for the star formation (SF) rate:
\begin{equation}
\Psi(t)=\nu M_g(t),
\end{equation}
where $\nu$ is a normalization parameter (SF efficiency) and $M_g(t)$ is
the residual mass of gas at any given galactic time.  During phases of active
star-formation,
stellar emission is assumed to be extinguished by dust in a uniform screen, 
modelled according to the standard extinction law ($R_V=A_V/E[B-V]=3.1$,
Cardelli et al. 1989).
A further relevant parameter is the
timescale $t_{infall}$ for the infall of primordial gas. The evolution
patterns for the different galactic populations considered here are obtained 
with the following choices of the parameters.

For early-type galaxies, we have set a quick infall timescale 
$t_{infall}=0.1~Gyr$ and a high SF efficiency, $\nu=2~Gyr^{-1}$.
The corresponding SF law has a maximum at a
galactic age of 0.3 Gyr, and is truncated at
0.8 Gyr to mimic the onset of a galactic wind. During this quick
star-forming phase, the galaxy emission is assumed to be
extinguished by $A_V=6$ magnitudes.

For late-type galaxies, we adopted a longer $t_{infall}=4~Gyr$
and a correspondingly lower efficiency $\nu=0.6~Gyr^{-1}$. In this case,
the peak of the SF occurs at 3 Gyr and a galactic wind is never produced.
The same parameters assumed to reproduce the spectra of spirals
and irregular galaxies.
This may not be entirely representative of a galaxy
during a starburst phase, but, given the spectral region considered
in this work, our assumption is still a good approximation.
We have then generated two grids of model spectra for both early- and late-types
spanning a range of galactic ages from 0.1 to 15 Gyr.

Our assumed local luminosity functions (LLF) at 3.6 $\mu$m and in 
the $K$-band have been derived from those estimated by Kochanek et al. 
(2001) for both the early-type and late-type galaxy classes, based on a 
K-band selected sample taken from the Two Micron All Sky Survey (2MASS)
and including 4192 low-redshift ($z \sim 0.02$) galaxies.
We have made use of their luminosity functions differentiated by
morphological type according to the de Vaucouleurs's parameter 
$T$ estimated by the authors, and adopting the value $T=-0.5$ 
as the boundary between spheroidal and late-type galaxies.
[For spheroids the $K$-band Schechter best-fit parameters
are $\alpha=-0.92$, $M^{\ast}_K = -23.53$ and
normalization factor $n^{\ast}=0.45\times 10^{-2}\ Mpc^{-3}$ 
for $H_0 =100 \ km \ sec^{-1}  Mpc^{-1}$].
Transformation from 2.2 to 3.6 $\mu$m is performed with the SED templates
for the two classes at the present cosmic time.

In our schematic evolutionary model, we have assumed that, once formed
at a given redshift, the comoving number densities of the spiral population 
keeps constant, while the galaxy luminosities evolve following their 
evolutionary stellar content. 
This choice reflects our assumption that, once having acquired its final
morphological structure within the Hubble sequence, a normal galaxy 
evolves only due to the secular change of the integrated stellar spectrum.

For the spiral galaxy class we assumed a high redshift of formation ($z_{form}=5$,
but the specific value is by no means critical, any other choice
between 2 and 5 would give essentially the same results)
and constant number density henceforth. Spheroidal models are developed
in Sect. \ref{ES0} below.

\subsubsection{A population of fast evolving starbursts}
\label{SB}

We have seen in Sec. \ref{vvmax} that the star-forming galaxies 
(the highly asymmetric, $A>0.4$, mergers/irregulars) display 
a $V/V_{max}$ distribution indicative of an excess of 
sources at high-redshifts.
This, as well as number counts and z-distributions, are
not consistent with the assumption that the galaxy luminosity 
function evolves purely following the secular evolution of the constituent 
stellar populations.

The inability to reproduce the faint galaxy counts in the B-band with no-evolution
prescriptions is also a well-established result (Ellis 1997). 
Even more evident departures from no-evolution were reported
from deep observations in the mid- and far-IR (e.g. Franceschini et al. 2001; 
Elbaz et al. 2002; Gruppioni et al. 2002; Lagache, Dole \& Puget 2003).                 
All this indicates the presence of a numerous population of 
irregular/merging systems at high-redshifts, likely suggesting 
luminosity as well as density evolution going back in cosmic time. 

We then added to our modellistic description a 
population of starburst galaxies whose comoving number density $\rho(z)$ 
evolves according to:
\begin{equation}
\rho(z) \propto \rho(z_0) \times (1+z) 
\end{equation}
for $z < 1$, keeping constant above, and whose luminosities $L(z)$ also increase as
\begin{equation}
L(z) = L(z=0) \times exp[k\cdot \tau(z)] ,
\end{equation}
where $\tau(z)=1-t_H(z)/t_H(z=0)$ is the look-back time in units of the
present Hubble time $t_H(z=0)$, and the evolution constant is $k=1.7$ for $0<z<2$,
and $k=0$ at $z\geq 2$ 
[such that $L(z=1)\simeq 2.6\cdot L(z=0)$ and $L(z=2)\simeq 4\cdot L(z=0)$]. 

For the local luminosity function of this
population of irregular/merging starbursts, we have referred to that
obtained in the B-band by Franceschini et al. (1988),
based on a local, morphologically selected, sample (from UGC), with complete
spectroscopic identification.
We have both appropriately transformed this LLF to the $K$-band and taken care to 
slightly rescaling it in such a way that the sum of this with the LLF of spiral 
galaxies would match the $K$-band LLF by Kochanek et al. (2001).

\subsubsection{An empirical evolutionary schemes for spheroidal galaxies}
\label{ES0}

We have considered two simplified schemes of the formation of 
spheroidal galaxies for immediate comparison with the data,
both of them tied to fit the LLF derived from Kochanek et al. (2001).
The first one is a classic prescription assuming a single
impulsive episode for the formation/assembly of the field ellipticals, 
occurring at a high redshift 
($z_{form} > 2.5$, e.g. Daddi et al. 2000 and Cimatti et al. 2002b),
and Pure Luminosity Evolution (PLE) thereafter.
We assumed a redshift of formation $z_{form}=3.0$.
In this case, the birth of stars coincides with the formation of the spheroid.
In the following we will refer to this as the $monolithic$ formation model.

The second model for spheroids describes a situation 
in which massive ellipticals form (or at least assemble) at lower redshifts 
through the merging of smaller units down to recent epochs. In such a case 
their formation is not a single coeval process, but
is spread in cosmic time.
We achieved this by splitting the local spheroidal galaxies into 
several sub-populations, each one forming at different redshifts. For simplicity,
we assumed that all sub-populations have the same mass and luminosity functions 
and differ only for the normalizations, whose total at $z=0$ has to 
reproduce the local observed luminosity function.
We emphasize that this assumption of a luminosity function building up with
time by keeping a constant shape may result as oversimplified in the
light of our results in Sect. \ref{mf}. However we consider it as a
useful reference and defer more sophisticated treatment to future papers.

We have calibrated and tested this model against deep galaxy surveys in the K band
(K20: Cimatti et al. 2002a,2002b, HDF's: Franceschini et al. 1998 and
Rodighiero et al. 2001; GDDS: Abraham et al. 2004).
In our current implementation, we assume 7 spheroidal sub-populations, the 
bulk of which ($\sim 80\%$) form in the redshift 
interval $0.9 < z < 1.6$, with additional fractions being produced at
higher (starting from $z\sim 5$) and lower $z$ (down to $z\sim 0.5$). 
The detailed fractions of field ellipticals as a function of their formation
redshift in our model is summarized in Table \ref{tab:best-fit}.

Of this seemingly $hierarchical$ scheme, that we name 
as the $Protracted$-$Assembly$ (PA) model, we have considered two applications.
(The $PA$ terminology is intended to underline that our
scheme, although generically in line with the hierarchical expectations,
does not provide a physical discription, and at the same time is more general).

In our first considered case (hereafter $PA~model~1$) stars are assumed
to be coeval to the spheroid sub-populations forming at different redshifts
$z_{form}$.

We have also considered a different situation in which the structural 
assembly of stars in the host galaxies happens later than
their formation epoch. We have then modified our previous 
$Protracted$-$Assembly$ scheme by assuming two different epochs, one for the 
birth of stars and the other for their assembly.
In this second case (hereafter $PA~model~2$) all the stars
present in today ellipticals are assumed to be born at high redshift ($z =5$), 
while they are dynamically assembled in the various spheroid sub-populations
at the redshifts $z_{form}$ in Table \ref{tab:best-fit}.

In Fig. \ref{cint36} we compare the observed IRAC/GOODS 3.6 $\mu$m 
counts with the predictions of the three models discussed above:
our $PA~1$ (solid blue lines), the $PA~2$ (three 
dots-dashed red line) and the $monolithic$ model (dashed green lines).
These predictions differ only for their treatment of the
early-type population. We note a generally good agreement of both
$Protracted$-$Assembly$ predictions for spheroids, while the $monolithic$ model does
overpredict the elliptical number counts fainter than $S_{3.6} \sim
60\ \mu$Jy. This excess, by a factor $>$3, is very significant all the
way down to $S_{3.6} \sim 1\ \mu$Jy.
On the other hand, within the uncertaintes we cannot discriminate
between the two $Protracted$-$Assembly$ solutions.

Also the source counts for late-types are well reproduced, with a slight
tendency to underpredict them in the flux range
$40\ \mu Jy<S_{3.6}<100\ \mu Jy$.
In general, only the two solutions including the $PA$ 
description for spheroids reproduce well
the total source counts (Fig. \ref{cint36}, bottom panel).

Similar conclusions can be derived from the redshift distributions.
In Fig. \ref{Dz36} the observed redshift distribution of the 3.6 $\mu$m 
IRAC/GOODS sample with $S_{3.6}>$10 $\mu$Jy is compared with our models.
Again, the $PA$ predictions fit much better the observed
distribution of spheroids. 
$PA~model~1$ tends to slightly overpredict the number of sources 
above $z\sim 1.2$ (both for early- and late-type galaxies), while
this excess is not present in $model~ 2$
(because in this case all stellar populations are somewhat
older and dimmer at the epoch of their assembly into the galaxy).

The predictions of $PA~model~1$ for the $K$-band statistics are
presented in Figures \ref{cintK}-\ref{dzK20} for source counts and
redshift distributions, showing a generally good agreement with the 
observations.

\begin{table}
\begin{center}
\begin{tabular}{c c}
\hline
$z_{form}$ & $spheroidal$ \\
$~$ & $fraction$ \\
\hline
$z_1 =5.10$ &  $5\%$\\
$z_2 =3.38$ &  $5\%$\\
$z_3 =1.61$ &  $20\%$\\
$z_4 =1.37$ &  $20\%$\\
$z_5 =1.13$ &  $20\%$\\
$z_6 =0.89$ &  $20\%$\\
$z_7 =0.65$ &  $10\%$\\
\hline
\end{tabular}
\end{center}
\caption{Fractional contributions to the local mass function for various spheroidal 
galaxy sub-classes being formed at $z_{form}$, for our best-fit 
$Protracted$-$Assembly$ model.
}
\label{tab:best-fit}
\end{table}

\subsection{Comparison with Other Models }

We have compared our results with an improved version of the 
phenomenological evolution model by Xu et al. (2003). 
The model reproduces the bright end of the number counts, but 
starts to exceed the observations below $S_{3.6}<$25 $\mu$Jy,
for both spheroidal and late-type galaxies.

%
%
%

Silva et al. (2005) have elaborated a more physical model considering 
the mutual feedback between star-forming regions in galaxy spheroids and the 
active nuclei growing in their centers. 
This model's predictions for passive spheroids are in good agreement with 
the 3.6 $\mu$m IRAC/GOODS number counts at faint fluxes, but the model 
foresees too many
spheroidal galaxies at $S_{3.6}>$100 $\mu$Jy (by factors $\sim3-5$)
and too few spirals/irregulars (again by factors $\sim3$).
The model could perhaps be 
made more consistent with our data in case that the bulk of the 
high-z formed spheroids could be incorporated into massive spiral 
galaxies as their bulge components (hence escaping classification
as E/S0's at low-z).


\subsubsection{Predictions of the GALICS semi-analytic code}

GALICS is a hybrid model of hierarchical galaxy formation 
combining large cosmological N-body 
simulations with simple semi-analytic recipes to describe the fate of 
the baryons within dark matter haloes (Hatton et al. 2003). 

The available set of different simulated cones of universe allows one to check 
the effects of clustering and cosmic variance when dealing with observations
covering limited sky areas. We have used the GALICS 
database{\footnote {http://galics.cosmologie.fr/}} to simulate 10 cones 
of the universe, each covering an area comparable to the IRAC/GOODS survey.
The comparison of GALICS predictions with the 3.6 $\mu$m observed redshift 
distribution is reported in Fig. \ref{Dz36} (hatched grey regions).
A general agreement is observed for the total and also for the separate
early- and late-type populations. As in the case of our $Protracted-Assembly$
model 1 (see Sect. \ref{ourmodel}), the number of sources at $z>1.3$ 
is just slightly overpredicted.

\begin{figure}
\centerline{
\psfig{file=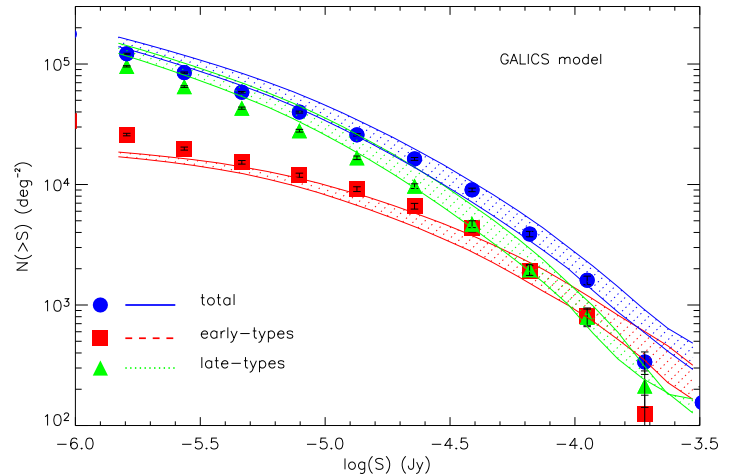,width=10cm}}
\caption{The 3.6 $\mu$m cumulative number counts from the
IRAC/GOODS sample corrected for incompleteness are compared with
the predictions of GALICS, a semi-analytical code(Hatton et al. 
2003).  Different symbols correspond
to different morphological classes (green triangles for late-types,
red squares for ellipticals, blue circles for the total). The model
report the contribution of different populations contributing to the
near-IR emission (green hatched region for late-types, red hatechd region
for spheroids and blue hatched region for the total). The width of
the hatched  regions shows the effects of cosmic variance over an area 
of 1 square degree. 
}
\label{cintgal}
\end{figure}

The observed number counts are compared in Figure \ref{cintgal} with 
the predicted effect of the cosmic variance shown as the hatched regions.
GALICS predictions look consistent with the total 
number counts, with a slight excess at $S_{3.6}<$10 $\mu$Jy
and a dearth of spheroidal galaxies at $S_{3.6}<$40 $\mu$Jy.
Within the uncertainties in the identification of the various morphological 
categories in the semi-analytic mock catalogues, this hierarchical 
code performs well in explaining the observational data.

%
%
%
%

\section{DISCUSSION}
\label{discussion}

\subsection{Phenomenological Models vs. Number Counts and z-Distributions
Data}\label{s1}

We have tried simple modellistic representations of the data on number 
counts and redshift distributions, to assist
our interpretation.   The most challenging constraints
concern the spheroidal galaxy population
and the modelling of their expected very luminous early phases.
Results of the match of modellistic expectations and observational 
data are briefly discussed in the following. 

\subsubsection{A Zeroth-order Solution: the Monolithic PLE Model}
\label{sol1}

The simplest evolutionary scheme that we have used for comparison
with the data is a monolithic PLE model.
Once the local LF is established by low-$z$ near-IR surveys 
(Sect. \ref{ourmodel}), this crude representation has quite few 
parameters to play with, and to correct for the inconsistencies that it
revealed in the previous Sect. Apart from the formation epoch ($z_{form}$ 
should in any case be $\sim3$ or larger), the most important model
parameters are those ruling the stellar IMF.
In our application we considered the standard Salpeter form with
differential spectral index $x=1.35$, ($N[M]\propto M^{-x}$):
with this ingredient, the expected number of young spheroidal galaxies 
between $z\sim1$ and 2.5 exceed the observations (see top left 
panel of Fig. \ref{Dz36}). 

The effects of a different choice for the stellar IMF and of 
variations in $z_{form}$ were discussed
in some details by Kitzbichler and White (2004), who compared 
(monolithic) PLE models with number count, z-distributions
and the galaxy M/L ratios from $K$-band surveys. They found similar 
inconsistencies to our above and that, to cure them, one would 
need to adopt an IMF almost completely deprived of massive stars 
(with an IMF differential spectral index $x>2$).

We have tested the effect of a slighter modification of the IMF in our
monolithic model by changing from the Salpeter to the Scalo one (Scalo
1986), but the effect turned out to be a marginal reduction (by few
tens percent) of the excess number counts and z-distributions.  On the
other hand, a more radical change, like that of bringing the IMF
spectral index $x$ to values in excess of 2, would entail problematic
side effects. In particular, if we keep normalization to the local
census of stellar populations in normal galaxies, a very steep IMF
would prevent young galaxies to produce the metals observed in the
galaxy intra-cluster plasmas (Mushotzky \& Loewenstein, 1997;
Baumgartner et al. 2005) and would make impossible to explain the
observed IR and optical background light (which would perhaps rather
require a shallower IMF, $x<1.35$, richer in massive stars, see Madau
and Pozzetti, 2000; Franceschini et al. 2001).

An alternative might perhaps be to assume that the whole early phases 
of the spheroidal galaxy evolution happen inside a heavily dust
extinguished medium (e.g. Franceschini et al. 1994; Kitzbichler and 
White 2004; Silva et al. 2005). Such extinction should be very high
(several optical magnitudes) to prevent detection by the IRAC 3.6 
$\mu$m band. The main 
difficulty with this solution, however, stems from the request
by the observational constraints that such luminous 
galaxies should be kept enshrouded by a thick dust envelope
during their whole early life, i.e. several Gyrs -- corresponding 
to the $z$ interval from $z\geq3$ to $z\sim1$.  
Considering the modest average dust attenuation in low-z 
galaxies ($A_V \simeq 0.2-0.3$ magnitudes, Kauffmann et al. 2003),
and the results of hydrodynamical simulations of galaxy 
mergers (e.g. Mihos and Hernquist 1994), it is clear that 
such high obscuration only characterizes transient short-lived
evolutionary phases with violent redistribution of the dusty ISM,
on timescales of $\sim 0.1$ Gyr, like those
inferred for the ultra-luminous IR galaxies (Genzel et al. 1998;
Rigopoulou et al. 1999).

\subsubsection{An Improved Scheme:  {\sl Protracted-Assembly} for
Spheroidal Galaxies}
\label{sol2}

We have obtained much easier fits to the data with the alternative approach of
assuming a progressive build up of the spheroidal galaxies taking place over a 
significant fraction of the Hubble time.  We have empirically represented this by
splitting the luminosity function of spheroids into various sub-components
and by attributing to each one a formation redshift $z_{form}$ and a weight
(see Table \ref{tab:best-fit}).
For simplicity, the luminosity functions of the various sub-components were
assumed to keep the same shape. 
Then the observational constraints are matched by a solution where $\sim80\%$ 
of the final spheroidal mass function is assembled between $z\sim1.6$ 
and $z\sim0.9$ for a standard cosmology.

We have considered two specific implementations of this $Protracted$-$Assembly$ 
scheme: one based on the assumption that stellar populations are formed at
the same time of the whole galaxy formation, $z_{form}$. The 
alternative was that stars preexisted the galaxy assembly, e.g. 
being produced at $z\sim 5$ and assembled into 
galaxies only later at $z_{form}$. 
This latter assumption of a progressive build-up of already aged 
stellar populations seems to be slightly favored by the observations 
(Figs. \ref{cint36} and \ref{Dz36}).

In either case, the adoption of a $Protracted$-$Assembly$
in cosmic time for spheroidal galaxies (i.e. the essential 
postulate of hierarchical models) appeared to overcome most of the 
apparent inconsistencies of the monolithic model.

The $PA$ model is also reasonably consistent with the scanty existing data on
high-$z$ ($z>1.4$) spheroidal galaxies. Daddi et al. (2005a) have recently
color selected a sample in the Hubble UDF to $K\simeq 21$ and used the
ACS grisms for spectroscopic identification. The number of spheroidal
galaxies with $1.4<z<2.5$ over a 12 sq. arcmin area predicted by the $PA$
model is perfectly consistent with the 7 objects found by Daddi et al.

Labbe' et al. (2005) combined ultra-deep K-band and ISAAC imaging of 5 sq. 
arcmin in HDF-South to identify 3 passive spheroidal galaxies with 
$1.9<z<3.8$ and $K<22.5$ (though one of the three could host an AGN). 
The prediction of the $PA$ model at such high-$z$ depends
on the detailed parameters describing the small fraction of objects
forming at $z>2$ in Table \ref{tab:best-fit}: our best-fit predicts 1.5 
galaxies for the Labbe' et al. selection function.



\subsection{Evolutionary Trends for Luminosity and Mass Functions}
\label{trends}

\subsubsection{Evolution of the Global Mass Function}

Our investigation of the evolution of high-redshift galaxies followed 
two main routes. The first one was to infer general constraints from 
statistical observables like the number counts, $z$-distributions, 
and $V/V_{max}$ analyses, as discussed in the previous Sect. \ref{s1}.
The second approach was to derive direct information about the evolution
of the main galaxy distribution functions from our reference IRAC/GOODS
flux-limited sample, whose results are summarized in Fig. \ref{massfunc}.

\begin{figure}
\centerline{
\psfig{file=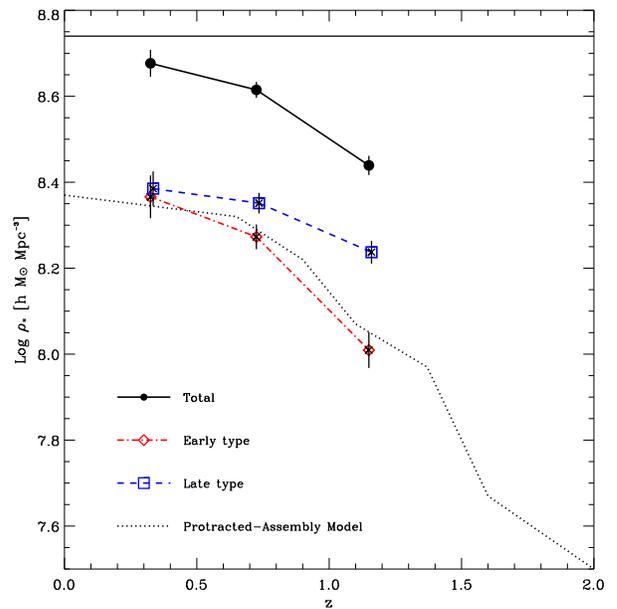,width=8.5cm}
}
\caption{Comoving integrated stellar mass density as a function of redshift,
split by morphology and integrated for $M_* h^2 >10^{10 }M_{\odot}$.
The solid horizontal line marks the local stellar mass density as
measured by Cole et al. (2001) over the same mass range.
This figure corresponds to a simple modification of fig. 8 in Bundy et al.
(2005), which was calculated with a higher mass cutoff ($M h^2 >10^{11}\ M_\odot$): 
in their case, the data indicated essentially no evolution of the comoving mass 
density from $z\sim 0$ to $z=1$,  while in our case the evolution is 
appreciable ($\sim 30-40$\% of mass decrease over the same $z$-interval).
The dotted line is the prediction of our $Protracted-Assembly$ model.
}
\label{zint}
\end{figure}

Do these independent lines of investigation bring to consistent solutions?
We have attempted to answer this question in Figure \ref{zint},
where we report the integrated comoving stellar mass density
as a function of the redshift for the two morphological classes
and for the total population. The integral has been computed using
galaxies with masses larger than $10^{10} h^{-2} \ M_\odot$, 
to ensure completeness within all three redshift bins. We find that
the mass density increases by $\sim50\%$ from z$\sim1.2$ to $z=0.3$
for the total sample.

The black dotted line in Fig.\ref{zint} shows the evolution of the 
integrated comoving mass density in spheroidal galaxies with 
$M h^2 >10^{10}M_\odot$, as predicted by our $Protracted$-$Assembly$ 
model (and derived from the distribution of formation epochs of the 
sub-populations in Table \ref{tab:best-fit}).
Indeed, this predicted redshift dependence matches reasonably well our direct 
determination of the evolutionary comoving density (red dot-dash line), 
based on the integral of the observational mass function. 

In essence, this good match provides consistent evidence that significant evolution 
of the  global mass function has occurred typically at $z\simeq 0.7$ to 2,  
also clearly indicated by the $V/V_{max}$ analyses in Sect. \ref{vvmax}.
    This is the most robust of our conclusions, and agrees with the results of
various other published analyses (Dickinson et al. 2003; Glazebrook et al. 
2004; Fontana et al. 2004). Our average rate of evolution of the total stellar
mass density $\rho_\ast$ can be approximated as an exponential fall-off with
redshift:
\begin{equation}
\rho_\ast (z) \sim 6.02\ 10^8\ h^3 exp\left[-{(2+z)^4 \over 141.6}\right]\ \ M_\odot/Mpc^3,
\end{equation}
while for the spheroidal population a good fit to data in Fig. \ref{zint}
is given by
\begin{equation}
\rho_\ast (z) \sim 2.79\ 10^8\ h^3 exp\left[-{(2+z)^5 \over 342.2}\right]\ \ M_\odot/Mpc^3.
\label{rho2}
\end{equation}

Our estimated decrease by an average factor $\sim$2.5 from $z=0$ to 1.2 
of the stellar mass density of E/S0 galaxies is then somewhat steeper 
than the 40\% decrease estimated by Treu et al. (2005a), based on previous work 
by Im et al. (2002). Also our observed functional dependence on redshift,
an exponential fall-off, is different from their assumed power-law
proportionality $\rho_\ast\propto(1+z)^{-0.6}$.

\begin{figure*}
\centerline{
\psfig{file=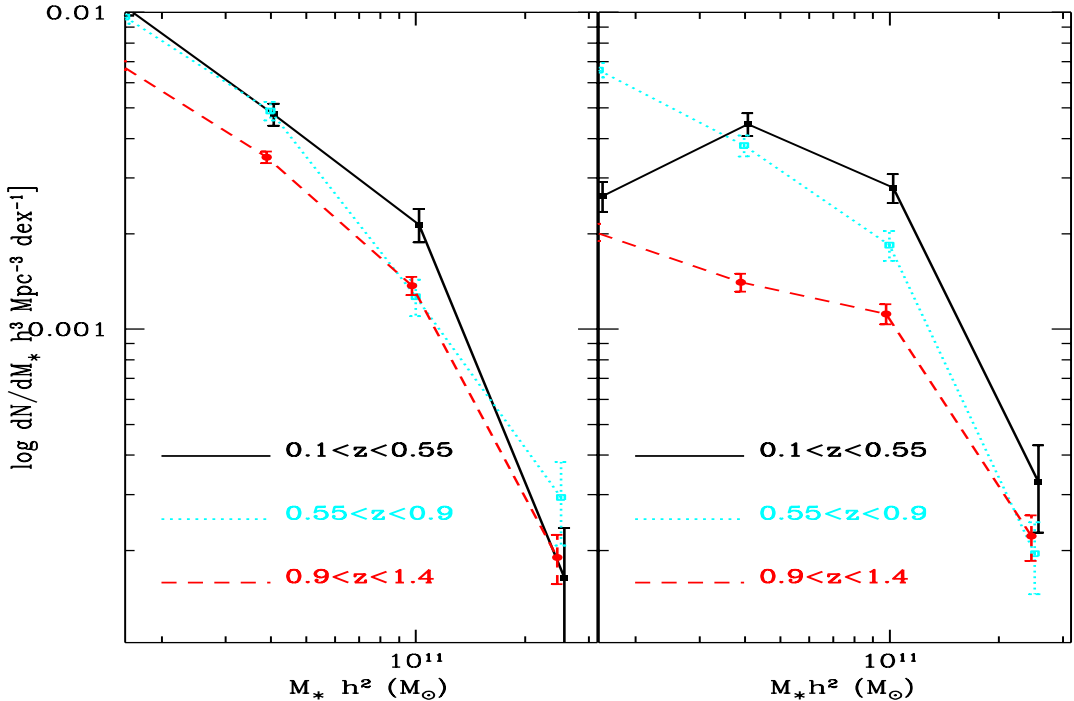,height=10cm,width=15cm}  }
\caption{The redshift-dependent mass functions for early-type ({\sl right-hand panel})
and late-type ({\sl left-hand panel}) galaxies in different redshift bins, as
indicated in the labels. Note that the function in the $0.55<z<0.9$ interval
might still suffer residual contamination by the galaxy concentrations
in that redshift bin (in spite that galaxies from $z=0.63$ to 0.77 are excluded).
(See also caption to Fig. \ref{massfunc}).
}
\label{mfunc}
\end{figure*}

%
%


\subsubsection{Mass-dependent and Morphology-dependent Effects}

Closer inspection of our direct mass and luminosity function 
determinations reveals, however, quite more complexity in the
evolution pattern than the simple overall decline with redshift
in Fig. \ref{zint}.

We report in Figure \ref{mfunc} a comparison of the evolutionary
mass functions at different redshifts for the two morphological categories.
First of all, the evolution of the mass functions in Figs. \ref{zint} 
and \ref{mfunc} shows a differential effect with morphological type.
For the late morphological types, the effect of a number density evolution
seems present but not large, while it is quite more significant 
for the spheroidal galaxy class (right panels in Fig. \ref{mfunc}).

An important aspect also revealed by Fig. \ref{mfunc} is that this decline is 
not uniformly shared by all galactic masses, but mostly concerns objects with 
$M h^2 \leq 10^{11}\ M_\odot$. The comoving number density of the most massive 
galaxies, instead, keeps remarkably stable from local to $z=1.2$, independently
if they belong to the spheroidal or the late-type category.  

The differential effect of the evolution rate with galaxy mass applies to
both morphological categories in the two panels of Fig. \ref{mfunc}: in 
both cases the highest mass galaxies show the lowest rate of evolution. 
Again, the differential effect is stronger for the spheroidal component.

This is in line with a similar result by Bundy et al. (2005), who 
find little evolution to z=1 for the highest galactic masses.  Indeed, 
Fig.\ref{zint} should be red together with fig. 8 of Bundy et al. (2005),
which shows the evolution of the same integral of the mass function,
above $M h^2 =10^{11}\ M_\odot$: in their case there is essentially
no evolution of the comoving mass density from $z=1$ to 0, while with our
lower stellar mass threshold the decrease is appreciable for both galaxy 
classes. The rate of galaxy assembly appears to be a strong function of the mass.
This is also in keeping with the low specific SFR estimated by Daddi et al. 
(2005b) in their most massive galaxies at $z>1.4$.

Treu et al. (2005a,b), based on a high-resolution spectroscopic study 
of a sample of field spheroidal galaxies in the GOODS-N area, have found 
significant evolution of the Fundamental Plane as a function of redshift and 
morphological properties, that can be explained as a change of the average
$M/L$ ratios. They also find that "this evolution depends 
significantly on the dynamical mass, being slower for larger masses".
It is interesting to note that their analysis is based on dynamical estimates
of the galactic masses.  
Similar conclusions are also reached by Juneau et al. (2005).
So this effect of differential evolution, a manifestation of the "downsizing"
process originally identified and investigated by Cowie et al. (1996), 
appears to respond to both the mass of the galaxy's 
stellar content, and to the total, possibly dark-matter dominated, mass.
Downsizing in the formation of stars in spheroidal galaxies
was also clearly indicated by Franceschini et al. (1998) from their analysis
of galaxies in the HDF-North (see also hints in Gavazzi et al. 1996).

\subsubsection{Evolution Patterns}

The synopsis of the evolution of the galaxy mass and luminosity functions
offered by Fig. \ref{massfunc} may shed some light onto the physical
processes driving them.
Let us start at low-z and progress in redshift, and let us first consider 
the evolution of the total mass function.
There is only one way to produce the number density evolution, which is
observed to become effective at $z>0.9$: this is via the intermediate-mass 
galaxies to decrease progressively in mass, hence migrating towards the left 
side of the figure. 
This was likely achieved through both merging of lower mass preexisting
objects and formation of new stars.   

That new stars are likely formed during the merging events accompaning the 
evolution  of the mass function
is shown by the luminosity evolution which is evident in the left-hand 
panels of Fig. \ref{massfunc}. 
In principle, (negative) number density evolution would be expected to
characterize the luminosity function in a similar way as it does for the
mass function in the right-hand panels. 
This negative density evolution appears as completely counter-balanced 
by an increase in luminosity, so that the net apparent effect is that of a 
(positive) luminosity evolution.  Clearly, if the average galaxy mass decreases 
and the average luminosity increases with redshift, this requires a steady 
decrease of the $M/L$ ratio.

Also of interest might be to try to differentiate the evolutionary paths among
the different morphological classes.  As for the spheroidal galaxies, two 
ways may produce in principle the observed strong number density evolution:
one is through the intermediate-mass objects increasing in mass, and 
migrating to higher masses at decreasing-$z$ (increasing cosmic time),
via merging acquisition and new star-formation. The other path is via 
morphological transformations of spirals/irregulars to relaxed
early-type morphologies, again consequence of merging episods or gas 
exhaustion.  All these processes are likely to happen together.

For the late-type galaxy population, and in spite of the apparent slow 
evolution of the mass function in Fig. \ref{mfunc}, the evolutionary pattern 
was likely very complex: on one side, late-types lose in favour of the spheroidal 
population due to morphological transformations, on the positive side they 
should increase by galaxy merging and star formation.

In conclusion, our analysis has confirmed that the cosmic epochs close to 
redshifts $\sim$1 to 2 have experienced an active phase of  star formation
at the low-mass end, and galaxy  transformations at the knee of mass function, 
$M h^2 \sim 10^{11} M_\odot$. Perhaps only the most massive galaxies in the 
field have been relatively quiet actors at this times, as shown by the 
corresponding slow evolution of their mass function.

\subsubsection{Matching with Independent Evidence of Evolution }

Although we do not have yet a complete understanding of the whole set of
phenomena, it is certainly not a chance that the redshift interval from 
$\sim 0.7$ to $\sim 2$ coincides with the cosmic epochs of peak activity 
revealed by IR searches of the dust-obscured star-formation (e.g. by ISO, 
Franceschini et al. 2001, Elbaz et al. 2002; and by Spitzer, Perez-Gonzales 
et al. 2005, Le Flock et al. 2005).

On this respect, it is also interesting to note a parallelism between the 
slow evolution to $z=1.4$ of our most massive galaxies and the history of
the star-formation in the most luminous IR galaxies, which are their
natural progenitors. Both the ULIRG samples analyzed with Spitzer by Perez-Gonzales 
et al. (2005) and Le Floc'h et al. (2005) and the very luminous SCUBA sources 
(e.g. Smail et al. 2002; Chapman et al. 2003) 
show a bolometric comoving emissivity peaking at high-$z$ 
($z>1.5$), whereas lower luminosity sources have peak emission at lower $z$.
From the tight relationship of the bolometric IR emission with the SFR
(e.g. Kennicutt 1998; Rowan-Robinson et al. 1997), the more luminous the IR 
galaxy is, the more accelerated its evolution and most confined backwards 
in cosmic time its main phases of star formation.
This meets, at least qualitatively, the requirements set by our analysis
of the mass and luminosity functions.

Finally, and in the light of the close relationship of quasar activity and
the host galaxy formation implied by the ubiquitous presence of 
super-massive black holes in the cores of all local massive galaxies,
it is remarkable to note the same dependence on source luminosity of the
cosmological evolution of the X-ray emissivity in a complete
unbiased sample of X-ray AGNs recently quantified by Hasinger, Miyaji 
\& Schmidt (2005). This similarity in the cosmic evolution of galaxies
and AGNs/quasars, and the luminosity-dependent effect, were already 
noticed by Franceschini et al. (1999).

A large amount of independent data seem to provide concordant evidence
for an evolution pattern in galaxies at $z\simeq 0.7$ to $\geq 2$
as a function of the object's mass (both stellar and total mass).  
These data, suggestive of a global 
"downsizing" in galaxy formation with cosmic time, require some important
correction factors (probably related to feedback by forming objects)
to the hierarchical dark matter assembly, which makes otherwise an
appropriate baseline scenario, as also confirmed here.

\section{CONCLUSIONS}\label{conclusion}

This paper is devoted to a systematic exploitation of public multi-wavelength 
data from the GOODS survey in the Chandra Deep Field South to
derive observational constraints on the emergence of the Hubble galaxy 
morphological sequence through cosmic time.
Critical data for this purpose are made available, in particular, by the 
very deep multicolor high-resolution imaging by HST/ACS and by 
the Spitzer Space Telescope deep photometric infrared imaging.
We also make use of extensive optical spectroscopic observations 
by the ESO VLT/FORS2 and VIMOS spectrographs.

Our main selection for faint high-redshift galaxies is based on deep
images by IRAC on Spitzer. We have selected from them a highly reliable 
IRAC 3.6$\mu$m sample of 1478 galaxies with $S_{3.6}\geq 10\ \mu$Jy for
detailed statistical analyses and for the derivation of mass and 
luminosity functions in bins of redshift.
We have also extended the morphologically-differentiated number counts 
down to a flux limit of $S_{3.6}= 1\ \mu$Jy.
We have carefully analyzed and thoroughly tested these data for 
completeness and reliability, based on simulations.

Forty-seven percent of the sample objects have spectroscopic redshift
from the VVDS, K20 and GOODS projects.   
For the remaining, we have used photometric redshifts from COMBO-17 
for galaxies below $z\sim 1$, while, for galaxies for which the COMBO-17 
guess was above 1, we have re-estimated the photometric redshifts with $Hyperz$. 
Deep K-band VLT/ISAAC imaging in the field is also used to derive further 
complementary statistical constraints and to assist the source identification 
and SED analysis.    

This very extensive dataset is then used to assess evolutionary effects in
the galaxy stellar mass and luminosity functions, while luminosity/density evolution 
is further constrained with the number counts and redshift distributions.
The estimate of galaxy stellar masses benefits in particular by the
constraint set by the IRAC 3.6 $\mu$m flux on the number of low-mass
stars.
The deep ACS imaging has allowed us to differentiate these evolutionary 
paths by morphological type, that our simulations show to be reliable 
at least up to $z\sim 1.5$.

The main results of the paper are hereby summarized.

\begin{itemize}

\item
We have derived luminosity functions at 3.6$\mu$m for various galaxy populations 
as a function of redshift up to $z=1.4$. After careful calibrations of the $M/L$ 
ratio, based on a detailed spectral fitting analysis to the observed SED's for
each sample galaxy, we have also obtained estimates of the evolutionary stellar
mass functions.
  On one side, the 3.6$\mu$m luminosity functions that we have derived 
show evidence for a positive, moderate luminosity evolution as a function of 
redshift (by $\sim0.7$mag in the L-band from $z=0$ to 1.2 for the most
massive galaxies, $M h^2 >10^{11}M_\odot$, in agreement with Treu et al. 2005a), 
likely due to stellar ages in galaxies becoming younger at increasing $z$. 
On the other hand, the corresponding global mass function shows evidence for 
an exponential decrease in the comoving density of galaxies 
($\rho_\ast\propto exp(-[2+z]^4/141.6)$) at the corresponding redshifts.

\item
      The galaxy number counts, z-distributions, the 
$\left<V/V_{max}\right>$ test, as well as our direct estimate of the stellar
mass function above $M_\ast h^2 =10^{10} M_\odot$, provide consistent evidence 
for a progressive dearth (by a factor $\sim2.5$ by $z=1.2$ for the stellar mass
density, see Fig. \ref{zint} and eq. \ref{rho2}) of the spheroidal
galaxy population to occur starting at low-$z$ and becoming quite 
significative at $z\geq 0.7$, paralleled by an increase in luminosity
(half a mag in L-band).   
Simple evolutionary models, fitting the fast convergence 
of the number counts and redshift distributions, and the evolutionary mass
function, require the main episodes for spheroidal build-up 
(of either old or newly-formed stellar populations) to happen 
between $z\sim 2$ and $z\leq 1$ for such field population, on average.

\item
This decrease in comoving density of galaxies with redshift shows, however,
a remarkable dependence on galaxy mass, being strong for moderate-mass, 
but almost absent until $z=1.4$ for high-mass galaxies, thus confirming 
previous evidence for a "downsizing" effect in galaxy formation
(e.g. Cowie et al. 1996; Franceschini et al. 1998).
By comparison with dynamical studies of the high-redshift spheroidal
population (Treu et al. 2005a,b), it is concluded that both stellar mass and 
total "dynamical" mass are  driving parameters of this differential
evolution.
This evolutionary pattern may also help explaining some inconsistencies 
in the evolution of galaxies at high redshifts previously reported by different
teams.
Our results appear consistent with recent reports by independent teams 
and selection functions (Fontana et al. 2004; Bundy et al. 2005).

\item
As for the complementary class of actively star-forming (irregular/merger) 
galaxies, deep Spitzer/IRAC 3.6$\mu$m and K-band observations
show them to evolve towards moderately higher luminosities and number densities 
up to $z\sim 1$ to 2, while normal spirals (those with asymmetry indices
$A<0.4$)  show similar, though slower, convergence at $z>1$ 
to that of spheroids.

\item
Our favored interpretation of the estimated mass functions and evolutionary 
trends for the two broad galaxy categories is that of a progressive
morphological transformation (due to gas exhaustion and, likely, merging) 
from the star-forming to the passively evolving phase starting
at $z\geq 2$ and keeping on down to $z\sim 0.7$. The rate of this process 
appears to depend on galaxy mass, being already largely concluded 
by $z\sim 1.4$ for the most massive systems.

\item
We finally discuss how well this evidence for a differential rate of galaxy 
build up with galactic mass compares with estimates of the SFR history 
based on deep far-IR surveys (e.g. Perez-Gonzales et al. 2005). 
A match between the two complementary views, of the history of SFR by
the best star-formation tracer (the bolometric flux) on one side,
and the rate of stellar mass build up traced by the near-IR emission 
on the other, would be achieved just by assuming that the progenitors of
the most massive galaxies are the most (bolometrically) luminous sources 
at high-$z$.  Ample evidence is accumulating in favor of the latter.

\end{itemize}

If the evolution pattern for galaxies to $z\sim 1.4$ is now close to be
understood, the knowledge of what exactly happened in the critical 
higher redshift era is still limited by very poor statistics in the number
of detected sources and lack of spectroscopic follow-up. 
How is the mass function behaving in detail at such high redshifts?
Does the "downsizing" trend continue there, as it might seem natural to
expect? 
In principle, the sensitivity of Spitzer/IRAC would allow accurate stellar 
mass determinations at these high-$z$, but quite further substantial effort 
with powerful spectrographs is needed before we get credible answers.
It is encouraging that much along this line has already been undertaken
(among others, by GDDS, Abraham et al. 2004, Juneau et al. 2005; 
FIRES, Franx et al. 2003; GMASS, Cimatti et al., in progress; COSMOS \& z-COSMOS, 
Scoville et al., Lilly et al., in progress).

\vspace{0.75cm} \par\noindent
{\bf ACKNOWLEDGMENTS} \par
\noindent
This work is based on observations made with the {\it Spitzer Space Telescope},
which is operated by the Jet Propulsion Laboratory, California Institute of
Technology under NASA contract 1407.
Support for this work, part of the Spitzer Space Telescope Legacy Science
Program, was provided by NASA through an award issued by the Jet Propulsion
Laboratory, California Institute of Technology under NASA contract 1407.

ACS was developed under NASA contract NAS 5-32865, and this research
has been supported by NASA grant NAG5-7697. We are grateful for an
equipment grant from  Sun Microsystems, Inc. The Space Telescope Science
Institute is operated by AURA Inc., under NASA contract NAS5-26555.

Many of the observations bringing to these results have been carried out 
using the Very Large Telescope at the ESO Paranal Observatory.

This work makes use the GalICS/MoMaF Database of Galaxies (http://galics.iap.fr).
We thank L. Silva for making available to us her code's results in tabular form
and she, Alvio Renzini and Andrea Cimatti for useful comments.
We warmly thank Laurence Tresse, the referee, for a careful reading of the
paper and numerous useful comments.

\begin{appendix}
\section{Numerical Values for the Luminosity and Mass Functions}\label{par:sources}

We report in Tables \ref{tab:lum} and \ref{tab:mass} numerical values of our
estimated luminosity and mass functions.
For ease of comparison with previous work, the values of the functions, luminosities
and masses are reported in terms of the parameter $h=H_0/100\ Km/s/Mpc$.

\begin{table*}
\begin{center}
\begin{tabular}{c c c c c c c c}
\hline
    z-interval & logL &  tot &  tot\_err &  early  &   early\_err & late  &    late\_err\\
\hline
$0.1<z<0.55$ & 8.5500 &  -1.7126 &  -2.9737  & -2.2738  &-3.2414  & -1.8521  & -3.0475\\
           & 8.8500 &  -1.8263 &  -3.1048  & -2.4829   &-3.4332  & -1.9344  & -3.1588\\
	   &   9.1500 &  -1.7782 &  -3.0808  & -2.2119   &-3.2977  & -1.9778  & -3.1806\\
	   &   9.4500 &  -2.1027 &  -3.2431  & -2.4038   &-3.3936  & -2.4038  & -3.3936\\
	   &   9.7500 &  -2.4829 &  -3.4332  & -2.7048   &-3.5441  & -2.8809  & -3.6322\\
	   &   10.0500&   -3.1819&   -3.7827 &  -3.1819  & -3.7827 & -12.0000 &     -\\

$0.55<z<0.9$  &  8.8500 &  -1.6367  & -3.0290 &  -2.1886  & -3.3444  & -1.7797  & -3.0868\\
	    &   9.1500 &  -1.8317  & -3.1850 &  -2.4337  & -3.4860  & -1.9566  & -3.2475\\
	    &   9.4500 &  -1.9506  & -3.2445 &  -2.2576  & -3.3980  & -2.2457  & -3.3920\\
	    &   9.7500 &  -2.3515  & -3.4449 &  -2.6678  & -3.6031  & -2.6378  & -3.5881\\
	    &   10.0500&   -3.1150 &  -3.8267&   -3.5129 &  -4.0256 &  -3.3368 &  -3.9376\\
    
$0.9<z<1.4$ &  8.8500 &  -3.3009 &  -4.1274 &  -3.6304  & -3.9926  & -3.5752  & -4.2433\\
	    &  9.1500 &  -1.8855 &  -3.3934 &  -2.5846  & -3.7524  & -1.9823  & -3.4394\\
	    &  9.4500 &  -2.2509 &  -3.6711 &  -2.8391  & -3.9674  & -2.3806  & -3.7352\\
	    &  9.7500 &  -2.4292 &  -3.7637 &  -2.8422  & -3.9702  & -2.6413  & -3.8697\\
	    &  10.0500&   -2.8824&   -3.9903&   -3.2598 &  -4.1790 &  -3.1185 &  -4.1083\\
	    &  10.3500&   -3.7717&   -4.4349&   -4.0727 &  -4.5854 &  -4.0727 &  -4.5854\\
\hline
\end{tabular}
\end{center}
\caption{Redshift-dependent luminosity functions for galaxies. 
Luminosities are $\log(L_{3.6\mum}*h^2/L_\odot)$. Luminosity functions are in units of
log(dN/dL) [$h^3*Mpc^{-3}*dex(L)^{-1}$], where $h=H_0/100\ Km/s/Mpc$.}
\label{tab:lum}
\end{table*}

\begin{table*}
\begin{center}
\begin{tabular}{c c c c c c c c}
\hline
z-interval &$log\ M/M_\odot$&   tot  &     tot\_err &  early  &   early\_err & late  &    late\_err\\
\hline
       
$0.1<z<0.55$&  9.0000  &  -1.6605  & -3.0617 &  -2.2536 &  -3.3592 &  -1.7885 &  -3.1163\\
           &   9.4000  &  -1.6934  & -3.1244 &  -2.1280 &  -3.2883 &  -1.8924 &  -3.2259\\
           &   9.8000  &  -1.8183  & -3.1678 &  -2.2461 &  -3.3684 &  -2.0213 &  -3.2657\\
           &  10.2000  &  -1.8914  & -3.2922 &  -2.5798 &  -3.5441 &  -1.9910 &  -3.3735\\
           &  10.6000  &  -2.0358  & -3.2721 &  -2.3526 &  -3.4305 &  -2.3216 &  -3.4150\\
           &  11.0000  &  -2.3068  & -3.4076 &  -2.5535 &  -3.5310 &  -2.6700 &  -3.5892\\
           &  11.4000  &  -3.3068  & -3.9076 &  -3.4829 &  -3.9957 &  -3.7840 &  -4.1462\\
       
$0.55<z<0.9$&   9.4000  &  -2.5541 &  -3.5767 &  -3.6378 &  -4.1506 &  -2.5915 &  -3.5676\\
            &   9.8000  &  -2.0710 &  -3.3666 &  -3.0336 &  -3.8462 &  -2.1211 &  -3.3914\\
            &  10.2000  &  -1.7891 &  -3.2470 &  -2.1817 &  -3.4648 &  -2.0146 &  -3.3440\\
            &  10.6000  &  -2.0610 &  -3.3616 &  -2.4203 &  -3.5418 &  -2.3107 &  -3.4858\\
            &  11.0000  &  -2.5075 &  -3.5854 &  -2.7347 &  -3.6990 &  -2.8975 &  -3.7804\\
            &  11.4000  &  -3.3368 &  -4.0000 &  -3.7089 &  -4.3011 &  -3.5317 &  -4.0625\\
       
$0.9<z<1.4$&   9.8000  &  -12.0000 & -        & -12.0000 & -        &  -2.8285 &  -3.9854\\
           &  10.2000  &   -2.0545 &  -3.5966 &  -2.6938 &  -3.8967 &  -2.1677 &  -3.6564\\
           &  10.6000  &   -2.3103 &  -3.7603 &  -2.8522 &  -4.0340 &  -2.4572 &  -3.8326\\
           &  11.0000  &   -2.6040 &  -3.9127 &  -2.9520 &  -4.0865 &  -2.8627 &  -4.0421\\
           &  11.4000  &   -3.3847 &  -4.3039 &  -3.6536 &  -4.4383 &  -3.7205 &  -4.4718\\

\hline
\end{tabular}
\end{center}
\caption{Redshift-dependent stellar mass functions for galaxies. 
Masses are $\log(M*h^2/M_\odot)$. The mass functions are in units of
log(dN/dM) [$h^3*Mpc^{-3}*dex(M)^{-1}$], where $h=H_0/100\ Km/s/Mpc$. }
\label{tab:mass}
\end{table*}

\end{appendix}

\end{document}